\documentclass[tightenlines,twoside,secnumarabic,
               onecolumn,floatfix,nofootinbib,preprint,11pt]{revtex4}

\usepackage{graphicx}
\usepackage{dcolumn}
\usepackage{float}
\usepackage{amsmath}
\usepackage{amsfonts}
\usepackage{multirow}
\usepackage{color}
 
\begin{document}
\def\nf{n_{\mskip-2mu f}}
\def\prop#1{{\cal P}_{#1}}
\def\tr{\mathop{\rm tr}\nolimits}
\def\Lsnew{\mathop{\rm \widetilde {Ls}}\nolimits}
\def\phid{\phi^\dagger}
\def\wh{\widehat}
\def\Ls{\mathrm{Ls}}
\def\Li{\mathrm{Li}}
\def\Ll{\mathrm{L}}
\def\I33m{\mathrm{I}_3^{3{\mathrm m}}}
\def\nn{\nonumber}
\def\fl{{\rm f}}
\def\be{\begin{equation}}
\def\ee{\end{equation}}
\def\bea{\begin{eqnarray}}
\def\eea{\end{eqnarray}}
\def\treenum{{(0)}}
\def\oneloopnum{{(1)}}
\def\nloopnum{{(n)}}
\def\qb{{\bar{q}}}
\def\cg{c_\Gamma}
\def\e{\epsilon}
\def\la{\langle}
\def\ra{\rangle}
\def\eps{\epsilon}
\def\ve{\varepsilon}
\def\musq{\mu^2}
\def\spa#1.#2{\left\langle#1#2\right\rangle}
\def\spb#1.#2{\left[#1#2\right]}
\def\lor#1.#2{\left(#1\,#2\right)}
\def\sand#1.#2.#3{%
\left\langle\smash{#1}{\vphantom1}^{-}\right|{#2}%
\left|\smash{#3}{\vphantom1}^{-}\right\rangle}
\def\sandp#1.#2.#3{%
\left\langle\smash{#1}{\vphantom1}^{-}\right|{#2}%
\left|\smash{#3}{\vphantom1}^{+}\right\rangle}
\def\sandpp#1.#2.#3{%
\left\langle\smash{#1}{\vphantom1}^{+}\right|{#2}%
\left|\smash{#3}{\vphantom1}^{+}\right\rangle}
\def\sandpm#1.#2.#3{%
\left\langle\smash{#1}{\vphantom1}^{+}\right|{#2}%
\left|\smash{#3}{\vphantom1}^{-}\right\rangle}
\def\sandmp#1.#2.#3{%
\left\langle\smash{#1}{\vphantom1}^{-}\right|{#2}%
\left|\smash{#3}{\vphantom1}^{+}\right\rangle}
\def\spab#1.#2.#3{\langle#1|#2|#3]}
\def\spba#1.#2.#3{[#1|#2|#3\rangle}
\def\g#1{#1_g}  
\def\Etmax{E_T^{\rm{max}}}
\def\fig{Fig.}
\def\tab{Table}
\def\alert#1{{\bf *** #1 ***}}

\title{Triphoton production at hadron colliders}

\author{John M. Campbell}
\email{johnmc@fnal.gov}
\affiliation{Fermilab, Batavia, IL 60510, USA}
\author{Ciaran Williams}
\email{ciaran@nbi.dk}
\affiliation{Niels Bohr International Academy and Discovery Center,
The Niels Bohr Institute, Blegdamsvej 17, DK-2100 Copenhagen \O, Denmark}

\preprint{FERMILAB-PUB-14-033-T}

\begin{abstract}
We present next-to-leading order predictions for the production of triphoton final states at the LHC
and the Tevatron.  Our results include the effect of photon fragmentation for the first time and we
are able to quantify the impact of different isolation prescriptions.
We find that calculations accounting for fragmentation effects at leading order, and those
employing a smooth cone isolation
where no fragmentation contribution is required, are in reasonable agreement with one another.
However, larger differences in the predicted rates
arise when higher order corrections to the fragmentation functions are included. 
In addition we present full analytic results for 
the $\gamma\gamma\gamma$ and $\gamma\gamma+$jet one-loop amplitudes. These amplitudes, which 
are particularly compact, may be useful to future higher-order calculations.
Our results are available in the Monte Carlo code MCFM. 
\end{abstract}
\maketitle

\section{Introduction}

The study of multiple photon production at hadron colliders has a long and rich history. Experimental analyses 
of isolated prompt diphotons have been undertaken for  nearly thirty years~\cite{Bonvin:1988yu,Albajar:1988im,Alitti:1992hn,Abe:1992cy}. 
More recent experimental studies at the Tevatron~\cite{Abachi:1996qz,Aaltonen:2011vk,Abazov:2010ah,Aaltonen:2012jd}
and the LHC~\cite{Chatrchyan:2011qt,Aad:2012tba,Aad:2013zba,Chatrchyan:2013mwa} 
have provided a wealth of precision data to compare with theoretical predictions.  At hadron colliders, prompt photons are primarily
produced through the ${\cal O}(\alpha_s \alpha)$ direct photon process, $h_1+h_2 \to \gamma + $jet.  Experimentally, this high-rate process
plays a special role in the calibration of jet energies and uncertainties by leveraging the simple kinematics of this process to relate the
well-measured photon to the less-understood recoiling jet.  The production of photon pairs, $h_1+h_2 \to \gamma \gamma$, occurs at a
much smaller rate due to the overall coupling involved, ${\cal O}(\alpha^2)$.  However, a detailed understanding of this channel is
particularly desirable in light of  the recent discovery of a Higgs boson~\cite{Aad:2012tfa,Chatrchyan:2012ufa} that decays
through the loop-induced process $H \to \gamma\gamma$.  With the accumulation of larger data-sets it should be possible to study more complex
final states that include additional hadronic jets, for instance the ${\cal O}(\alpha_s^2 \alpha)$ process $h_1+h_2 \to \gamma +$2~jets
or the ${\cal O}(\alpha_s \alpha^2)$ process $h_1+h_2 \to \gamma \gamma +$jet. Even the relatively rare triphoton process,
$h_1+h_2 \to \gamma \gamma \gamma$  should be accessible with existing data sets. Since such processes allow a much wider
range of kinematic regions, compared to simpler $2 \to 2$ reactions, one might expect their study to provide a more
thorough test of the theoretical predictions.

Experimentally, photons are identified as isolated -- i.e. they should be accompanied by little hadronic energy -- in order to distinguish
them from photons produced through other mechanisms, e.g. from neutral pion decays.  On the theoretical side it has become common
to treat the issue of isolation in one of two ways.  The traditional approach, which implements a parton-level equivalent of
an experimental isolation cut~\cite{Catani:2002ny}, requires the introduction of fragmentation functions that  describe the splitting of a parton into a photon.
These functions require non-perturbative input in a similar fashion to parton distribution functions (PDFs) and several sets are
available that have been tuned to data from the LEP experiments~\cite{Bourhis:1997yu,GehrmannDeRidder:1997gf}.
An alternative approach~\cite{Frixione:1998jh} has been advocated
which changes the isolation prescription in such a way that
fragmentation functions are not required.  This prescription, which has become known as ``smooth cone'' isolation,
thus enables a more straightforward calculation of higher-order theoretical predictions for photon processes.

Theoretical predictions for the production of  direct photons and photon pairs have been available at next-to-leading order (NLO) for 
some time~\cite{Binoth:1999qq,Catani:2002ny}.  More recently the NNLO corrections to the diphoton process have been computed using
the smooth cone prescription~\cite{Catani:2011qz}. Including the NNLO corrections increases the agreement 
between theory and data substantially, in particular for observables that are non-trivial for the first time at NLO,  
such as the azimuthal angle between the photons. 
Results for the production of diphotons plus one jet were considered in~\cite{DelDuca:2003uz}, using smooth cone isolation, and
extended to account for fragmentation effects in ref.~\cite{Gehrmann:2013aga}.
Smooth cone results for diphoton production in association with two jets, an important 
background for Higgs boson production through vector boson fusion, have also been 
presented recently~\cite{Gehrmann:2013bga,Bern:2013bha,Badger:2013ava,Bern:2014vza}.
 
In this paper we concentrate on the  $\gamma\gamma\gamma$ signature, i.e. triphoton production,
and use it to quantify the differences between the various isolation prescriptions.  Since, even at lowest order,
it contains three particles in the final state, it has a much richer kinematic structure than the simplest direct
photon and diphoton processes discussed above.  As a result we expect it to provide a clearer comparison of isolation effects.
Since triphoton production is suppressed by a power of the electromagnetic coupling $\alpha$ compared 
to the diphoton process, the rates are much smaller and, to date, no experimental analysis has observed 
this signature. Despite this, with typical LHC cuts, it has the largest cross section of the triple vector boson processes that are now beginning
to be probed at the LHC~\cite{ATL-PHYS-PUB-2013-006,CMS-PAS-SMP-13-009}.  

Next-to-leading order predictions, implementing smooth cone isolation, have been presented in
ref.~\cite{Bozzi:2011en}.  In this paper we will present a re-calculation of this process, using compact expressions for the
underlying matrix elements, and extend the previous treatment to allow for the inclusion of fragmentation effects.
For comparison we also present results for a similar process,  $\gamma\gamma+$jet production.
Although this final state is quite similar to triphoton production it allows us to investigate whether
the presence of a parton at leading order leads to qualitatively different behavior of the isolation algorithms.

This paper proceeds as follows.  A summary of the NLO calculations performed in order to produce the results
in this paper is given in section~\ref{sec:calc}, including compact analytic 
results for the $\gamma\gamma\gamma$ one-loop amplitude.  In section~\ref{sec:iso} we discuss the various forms of isolation 
employed in theoretical calculations and experimental analyses. In section~\ref{sec:isocomp} we 
present a comparison between the different isolation prescriptions, primarily for the case of triphoton production, but also
for $\gamma\gamma$+jet production. We study 
triphoton phenomenology for the LHC and the Tevatron in section~\ref{sec:results}.
Finally, we present our compact results for the $\gamma\gamma+$jet virtual amplitudes in the appendix.

\section{Calculation}
\label{sec:calc}

In this paper we present NLO calculations of the processes,
$p + p \to \gamma \gamma \gamma$ and $p + p \to \gamma \gamma+$jet. 
Although results for the one-loop 
virtual corrections to photon processes have previously been presented in ref.~\cite{DelDuca:1999pa}, in that case they were obtained by forming appropriate symmetric combinations of multiparton QCD
amplitudes such that gluons are effectively replaced by photons. Using this procedure one can use the $q\overline{q}ggg$ results
presented in ref.~\cite{Bern:1994fz} to obtain photon amplitudes.  
However, a numerical application of this procedure is both
inefficient and prone to additional numerical instability. For example, the singularities associated with non-Abelian diagrams
are not present in multiphoton amplitudes, but this is only made apparent through large numerical cancellations.
For this reason, we have re-computed the one-loop amplitudes using analytic unitarity
methods~\cite{Britto:2004nc,Britto:2006sj,Mastrolia:2009dr,Badger:2008cm}, and the program S@M~\cite{Maitre:2007jq},
in order to produce results that are as compact as possible.  We believe that these analytic formulae may be useful in the future,
for instance to optimize NNLO calculations of the diphoton process.

In this section we will present the one-loop amplitudes for the process,
\begin{equation}
0 \longrightarrow {\bar q}(p_1) + q(p_2) + \gamma(p_3) +\gamma(p_4) +\gamma(p_5) \;,
\end{equation}
where all momenta are outgoing and the momentum labels for the particles are given in parentheses.
The tree-level amplitude is written as, 
\begin{eqnarray} 
A^{(0)}(1_{\overline{q}}^{h_1},2_q^{h_2},3_{\gamma}^{h_3},4_{\gamma}^{h_4},5_{\gamma}^{h_5})
 ={i(\sqrt{2}\,e\,Q_i)^3}\mathcal{A}^{(0)}(1_{\overline{q}}^{h_1},2_q^{h_2},3_{\gamma}^{h_3},4_{\gamma}^{h_4},5_{\gamma}^{h_5})
\end{eqnarray}
where the helicities of the particles are denoted by $h_1, \ldots, h_5$.  Amplitudes with identical photon helicities vanish.  
As a result there is only one independent amplitude,
\begin{eqnarray} 
\mathcal{A}^{(0)}(1_{\overline{q}}^-,2_q^+,3_{\gamma}^+,4_{\gamma}^+,5_{\gamma}^-)
 = \frac{\spa1.2\spa1.5^2}{\spa1.3\spa1.4\spa2.3\spa2.4} \;,
\label{eq:lohel}
\end{eqnarray}
which corresponds to the maximally helicity violating (MHV) case.
The remaining helicity amplitudes can be obtained through conjugation and line-reversal symmetries. 

The one-loop amplitude can be decomposed as follows, 
\begin{eqnarray}
A^{(1)}(1_{\overline{q}}^{h_1},2_q^{h_2},3_{\gamma}^{h_3},4_{\gamma}^{h_4},5_{\gamma}^{h_5})
 =\frac{\alpha_s}{2\pi}\left(\frac{N_c^2-1}{N_c}\right){i(\sqrt{2}\,e\,Q_i)^3}
  \mathcal{A}^{(1)}(1_{\overline{q}}^{h_1},2_q^{h_2},3_{\gamma}^{h_3},4_{\gamma}^{h_4},5_{\gamma}^{h_5}) \;,
\end{eqnarray}
in terms of the  virtual MHV primitive amplitude which is given by, 
\begin{eqnarray}
&&\mathcal{A}^{(1)}(1_{q}^-,2_{\overline{q}}^+,3_{\gamma}^+,4_{\gamma}^+,5_{\gamma}^-)
=\bigg[-\frac{1}{\epsilon^2}\bigg(\frac{\mu^2}{-s_{12}}\bigg)^{\epsilon}
-\frac{3}{2\epsilon}\bigg(\frac{\mu^2}{-s_{25}}\bigg)^{\epsilon}
-3\bigg]\mathcal{A}^{(0)}(1_{q}^-,2_{\overline{q}}^+,3_{\gamma}^+,4_{\gamma}^+,5_{\gamma}^-)  \nonumber\\
&&+\frac{\spa1.3^3\spa2.4\spa4.5^2-\spa1.4^3\spa2.3\spa3.5^2}{\spa1.3\spa1.4\spa2.3\spa2.4\spa3.4^3}
   \Ls_{-1}\bigg(s_{12}; s_{35},s_{45}\bigg)
  -\frac{\spa1.2^2\spa4.5^2}{\spa1.3\spa2.4^3\spa3.4}\Ls_{-1}\bigg(s_{13};s_{45},s_{25}\bigg)\nonumber\\
&&+\frac{\spa1.2^2\spa3.5^2}{\spa1.4\spa2.3^3\spa3.4}\Ls_{-1}\bigg(s_{14};s_{35},s_{25}\bigg)
+\frac{\spa1.5^2}{\spa1.4\spa2.3\spa3.4}\Ls_{-1}\bigg(s_{23};s_{45},s_{15}\bigg)\nonumber\\
&&-\frac{\spa1.5^2}{\spa1.3\spa2.4\spa3.4}\Ls_{-1}\bigg(s_{24};s_{35},s_{15}\bigg)
-\frac{\spa1.2^2\spa3.5^2}{\spa1.3\spa2.3^2\spa2.4\spa3.4}\Ls_{-1}\bigg(s_{45};s_{13},s_{12}\bigg)\nonumber\\
&&+\frac{\spa1.2^2\spa4.5^2}{\spa1.4\spa2.4^2\spa2.3\spa3.4}\Ls_{-1}\bigg(s_{35};s_{14},s_{12}\bigg)
-\frac{\spa1.5^2}{\spa2.4\spa1.3\spa3.4}\Ls_{-1}\bigg(s_{35};s_{12},s_{24}\bigg)\nonumber\\
&&+\frac{\spa1.5^2}{\spa2.3\spa1.4\spa3.4}\Ls_{-1}\bigg(s_{45};s_{12},s_{23}\bigg)
-\frac{\spa1.2\spa2.5^2\spb3.2}{\spa2.3\spa2.4^2}\frac{L_0(-s_{13},-s_{45})}{s_{45}} \nonumber\\&&
-\frac{\spa1.2\spa2.5^2\spb4.2}{\spa2.4\spa2.3^2}\frac{L_0(-s_{14},-s_{35})}{s_{35}} 
+\frac{\spa1.2\spa4.5^2\spb4.3}{\spa2.4^2\spa3.4}\frac{L_0(-s_{13},-s_{25})}{s_{25}}\nonumber\\&&
+\frac{\spa1.3\spa4.5^2\spb4.3^2}{2\spa2.4\spa3.4}\frac{L_1(-s_{13},-s_{25})}{s_{25}^2}
+\frac{\spa1.2\spa3.5^2\spb4.3}{\spa2.3^2\spa3.4}\frac{L_0(-s_{14},-s_{25})}{s_{25}} \nonumber\\&&
-\frac{\spa1.4\spa3.5^2\spb4.3^2}{2\spa2.3\spa3.4}\frac{L_1(-s_{14},-s_{25})}{s_{25}^2}
-\frac{\spa1.2\spa1.5\spa2.5}{\spa1.3\spa2.3\spa2.4^2}\log{\bigg(\frac{s_{45}}{s_{25}}\bigg)}\nonumber\\&&
-\frac{\spa1.2\spa1.5\spa2.5}{\spa1.4\spa2.4\spa2.3^2}\log{\bigg(\frac{s_{35}}{s_{25}}\bigg)}
+\frac{\spb3.4}{2\spb2.5}\bigg[
 \frac{\spa1.5}{\spa2.5\spa3.4} \left( \frac{\spa3.5}{\spa2.3} + \frac{\spa4.5}{\spa2.4} \right)
 +\frac{1}{\spb1.5} \left( \frac{\spb2.3}{\spa2.4} - \frac{\spb2.4}{\spa2.3} \right)
 \bigg] \;.
\end{eqnarray}
The amplitude is written in terms of the integral functions $\Ls_{-1}$, $L_0$ and $L_1$ that are defined by, 
\begin{eqnarray} 
\Ls_{-1}(x; y,z) &=&{\rm{Li}}_2\left(1-\frac{y}{x}\right)+{\rm{Li}}_2\left(1-\frac{z}{x}\right)
+\log{\frac{y}{x}}\log{\frac{z}{x}}-\frac{\pi^2}{6} \\
L_0(x,y) &=& \frac{\log(x/y)}{1-x/y} \\
L_1(x,y) &=& \frac{L_0(x,y)+1}{1-x/y}.
\end{eqnarray}
The amplitudes for $\gamma\gamma+$jet production are presented in the appendix.

The contribution of real radiation diagrams is straightforward and compact results
have already been given  in ref.~\cite{DelDuca:1999pa}.  The amplitudes have been implemented
in the Monte Carlo program MCFM~\cite{Campbell:1999ah,Campbell:2011bn,MCFMweb}, which handles the cancellation of singularities using Catani-Seymour
dipole subtraction~\cite{Catani:1996vz}.  These calculations will be available in v6.8 of the MCFM code.
For the case of triphoton production we have checked the validity of our results by finding excellent
agreement with the smooth cone isolation result that may be obtained from the VBFNLO code~\cite{Bozzi:2011en}.
We defer our discussion of the comparison with existing results for diphoton+jet production to section~\ref{sec:isocomp}.

\section{Photon isolation and fragmentation} 
\label{sec:iso}

Experimental searches for prompt photons, those which participate in the hard scattering process,
are complicated by the presence of secondary photons and photons arising from fragmentation processes. 
Secondary photons are those resulting from the decays of unstable particles
(for instance $\pi^0\rightarrow \gamma\gamma$), 
whilst fragmentation photons are produced from the splitting of a QCD parton. Both of these 
types of photons are typically accompanied by hadronic energy and thus can be suppressed by the application of
isolation cuts. 

For this reason experimental analyses typically apply fairly strict isolation criteria to photon candidates.
The isolation region is defined by a cone of radius $R_0 = \sqrt{\Delta \phi^2 + \Delta \eta^2}$ around the photon,
where $\Delta\phi$ and $\Delta\eta$ refer to the difference in azimuthal angle and pseudorapidity from the photon
respectively.  One definition of the isolation requirement is to demand that the sum of the hadronic energy in the
transverse direction inside this cone is less than some fixed value $E_T^{\rm{max}}$, 
\begin{eqnarray}
\sum_{{\rm{had}} \in R_0}  E_T^{{\rm{had}}} < E_T^{\rm{max}}.
\label{eq:fixedE}
\end{eqnarray}
Throughout this paper, when such a cut is applied we will refer to the procedure
as ``fixed energy'' isolation.  At the LHC, typical values for $E_T^{\rm{max}}$ range from  $5$--$50$~GeV.

An alternative strategy is to
require that the total hadronic energy is less than some fixed fraction of the photon transverse momentum
$\epsilon_{\gamma}$, 
\begin{eqnarray}
\sum_{{\rm{had}} \in R_0}  E_T^{{\rm{had}}} < \epsilon_{\gamma} p_T^{\gamma}.
\label{eq:fracE}
\end{eqnarray}
This will be referred to as ``fractional energy'' isolation.
For analyses involving energetic photons such a prescription may be more desirable since high-$p_T$ photons
can be accepted even if they are accompanied by hadronic activity that exceeds a fixed threshold chosen
for more typical, softer photons.

Obtaining theoretical predictions for final states that include photons also requires particular care.
At LO a process involving a fixed number of photons and jets is rendered finite by the cuts needed to define
the final state objects, provided, for example, one defines a jet-photon separation minimum. 
However, at NLO matters are complicated by the collinear singularity associated with a quark-photon 
splitting.  The singularity 
cannot be removed in a theoretically safe manner by simply applying a parton-photon separation requirement, since this 
cut would remove a slice of soft gluon phase space and spoil the cancellation of infrared singularities. 
In order to produce 
a finite cross section one must absorb the collinear singularity into a fragmentation function, in an analogous 
manner to the mass factorization of the initial state collinear singularities into the PDFs.
In order to  estimate the non perturbative boundary conditions one must extract the fragmentation function from a fit to data.
We shall use fragmentation functions that have been obtained by fitting data from the LEP experiments, that
correspond to the results of Ref.~\cite{Bourhis:1997yu} (``BFG'') and Ref.~\cite{GehrmannDeRidder:1997gf} (``GdRG'').

An alternative procedure that does not require the introduction of fragmentation functions is
the isolation prescription of Frixione~\cite{Frixione:1998jh}, often referred to as ``smooth cone'' isolation.
This requires that the hadronic energy in the vicinity of the photon satisfies the following condition, 
\begin{eqnarray} 
\sum_{{\rm{had}}}  E_T^{{\rm{had}}}\theta(R-R_{{\rm had},\gamma})< \epsilon_{\gamma} p_T^{\gamma}
  \left(\frac{1-\cos{R}}{1-\cos{R_0}}\right)^{n} \qquad \mbox{for all}~R \leq R_0\;.
\label{eq:smooth}
\end{eqnarray}
Using this isolation prescription it is clear that the collinear pole is removed, but that arbitrarily soft emissions are retained,
thus preserving the required cancellation of singularities.
Given its simplicity this type of isolation is widely used in theoretical calculations. However, due to the discrete nature of the 
calorimeter cells in experimental detectors, this type of isolation is difficult to impose experimentally. Recently the possibility 
of combining the two approaches, by using a series of staggered cones, has been studied in ref.~\cite{Catani:2013oma}.

\section{Comparison between isolation procedures}
\label{sec:isocomp}

\subsection{Isolation effects in $\gamma\gamma\gamma$ production}

In this section we investigate the impact of the isolation prescription on predictions for triphoton production.
Specifically, we will compare predictions obtained using the fixed energy, fractional energy and smooth cone isolation
procedures that are defined by Eqs.~(\ref{eq:fixedE}),~(\ref{eq:fracE}) and~(\ref{eq:smooth}) respectively.
Throughout this paper we will use the customary choice $n=1$ in Eq.~(\ref{eq:smooth}).
For the sake of this comparison we compute NLO cross sections for the LHC operating at 14 TeV, using the default MCFM electroweak
parameters that correspond in particular to $\alpha=1/132.338$.  We use the CT10 PDF set~\cite{Lai:2010vv}
and set the renormalization, factorization and 
fragmentation scales to be the invariant mass of the photonic system $\mu=m_{\gamma\gamma\gamma}$.
The final state is defined by a basic set of cuts on the photons,  
\begin{eqnarray} 
p_T^{\gamma} > 30 \; {\rm{GeV}}, \quad |\eta_{\gamma}| < 2.5 \;, \quad R_{\gamma\gamma} > 0.4 \;. 
\label{eq:basiccuts}
\end{eqnarray}
For the fixed and fractional energy isolation procedures, the calculation also depends on the choice of fragmentation functions.
We consider three such sets here.  The first two sets, obtained by Gehrmann-de-Ridder and Glover (GdRG)~\cite{GehrmannDeRidder:1997gf},
correspond to strictly fixed-order extractions at $\mathcal{O}(\alpha)$ (LO) and $\mathcal{O}(\alpha\alpha_s)$ (NLO). The final
set (BFG) includes a resummation of ${\cal O}(\alpha_s^n\log^{n+1}{\mu_F^2})$ corrections and corresponds to set II of ref.~\cite{Bourhis:1997yu}. 

\begin{center}
\begin{figure} 
\includegraphics[width=8cm]{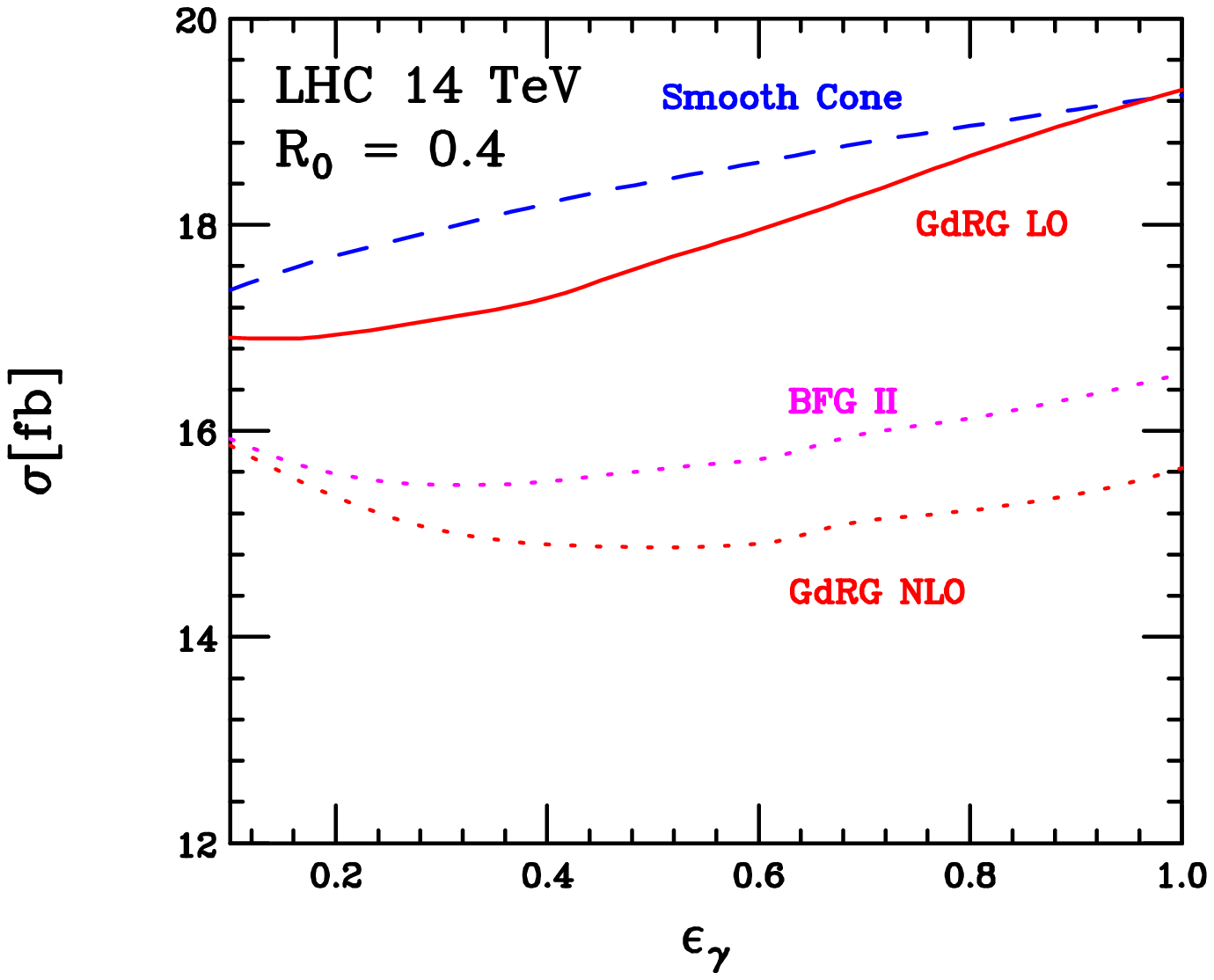}
\includegraphics[width=8cm]{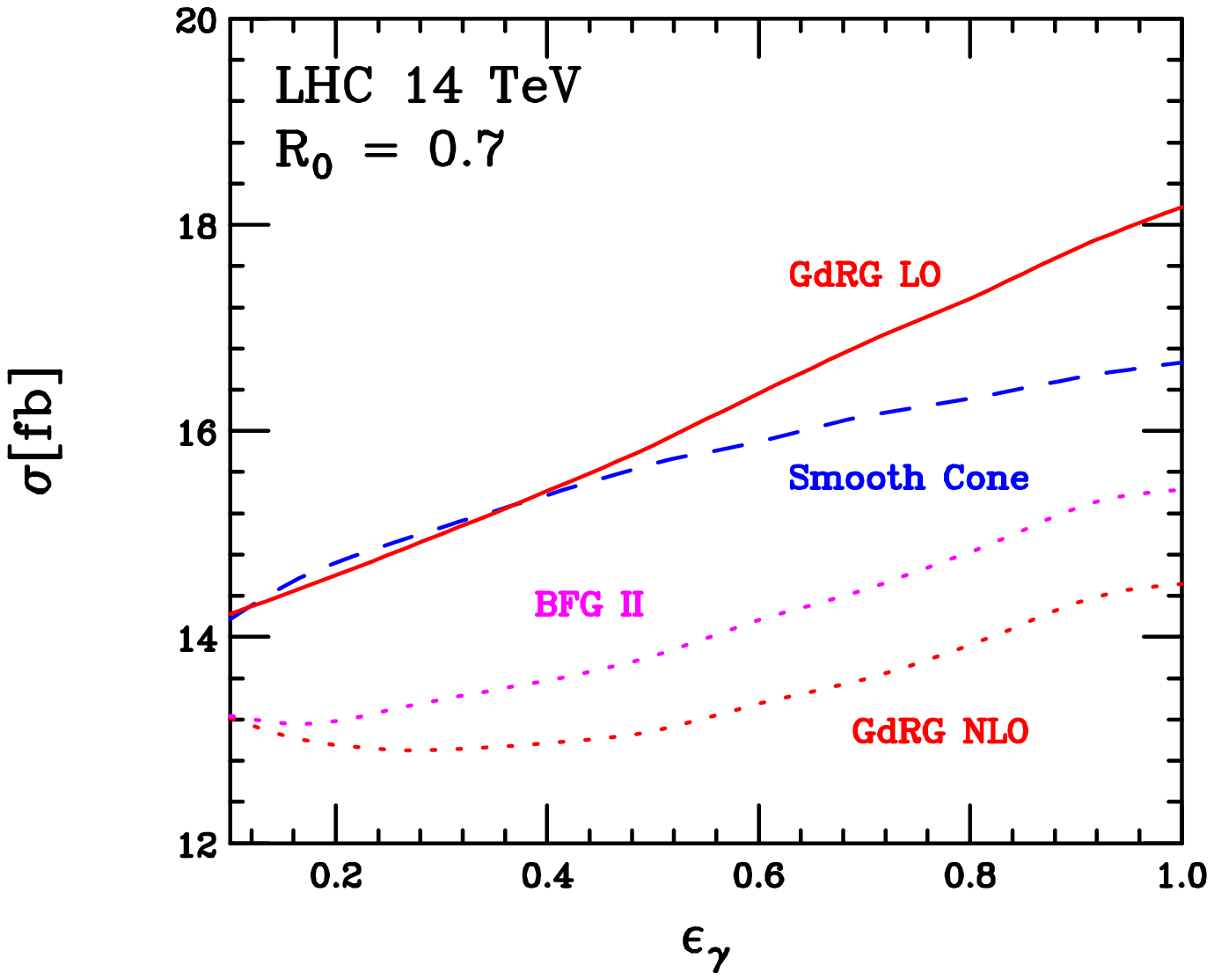}
\caption{Dependence of the NLO triphoton cross section on the parameter that controls the amount of hadronic energy inside
the isolation cone, $\epsilon_\gamma$.  Results are shown for the fractional and smooth cone isolation procedures, using
an isolation cone of size $R_0=0.4$ (left) and $R_0=0.7$ (right).
Smooth cone predictions correspond to the dashed line, while the solid line represents the LO GdRG prediction and the dotted lines correspond
to the BFG (magenta) and NLO GdRG (red) fragmentation sets.}
\label{fig:epscomp}
\end{figure} 
\end{center}
In Fig.~\ref{fig:epscomp} we compare the NLO cross sections for fractional and smooth cone isolation as a function of the parameter
$\epsilon_\gamma$ that is common to both algorithms.  We consider two different choices of isolation cone size, $R_0=0.4$ and $R_0=0.7$.
We first note that the LO prediction does not depend on the isolation procedure and, using the appropriate LO PDF set (CTEQ6L1), the
LO cross section is $6.90$~fb.   Regardless of the form of isolation used in the NLO calculation, the correction to the LO rate is
around a factor of two or more. However it is clear that the cross section is quite sensitive to  
the value of $\epsilon_{\gamma}$.  This sensitivity is 
easily understood from the nature of the final state. Since this process proceeds only through quark-antiquark initial states at tree level,
the effect of the NLO corrections is especially important due to the large gluon flux at the 14 TeV LHC.  Contributions of this nature,
for example real radiation channels such as $qg \to \gamma\gamma\gamma q$, are the most sensitive to the
the fragmentation functions and isolation definition due to the presence of a quark in the final state.

It is also clear from Fig.~\ref{fig:epscomp} that the predictions are rather sensitive to the fragmentation functions that are
employed.  The results for the LO GdRG set agree reasonably well with those using the smooth cone isolation.
For the smaller cone choice these two predictions differ by around 5\%, while for $R_0=0.7$ some differences at the 10\% level are observed
for the largest values of $\epsilon_{\gamma}$.  In contrast, the results obtained using the NLO set of GdRG and the BFG set II are
consistently $5$--$10$\% smaller than the results for LO GdRG.\footnote{ 
We note that the fitting range of the GdRG fragmentation sets corresponds to $\epsilon_{\gamma} \lesssim 0.5$ and that results
may not be reliable outside this range.  However, the GdRG and BFG fragmentation sets do not differ greatly in the region $\epsilon_{\gamma} > 0.5$.
Since the BFG sets use more inclusive LEP data, this similarity gives some confidence in the GdRG set in this region.}
We note though that the predictions obtained using these sets are less
sensitive to the isolation parameter $\epsilon_{\gamma}$ and the two sets yield very similar predictions for tightly isolated photons,
$\epsilon_{\gamma} \lesssim 0.2$.

Since, in our implementation, the QCD matrix elements which multiply the fragmentation contributions 
are $\mathcal{O}(\alpha^2\alpha_s)$, a consistent $\mathcal{O}(\alpha^3\alpha_s)$
prediction for triphoton production is only obtained when using the $\mathcal{O}(\alpha)$ LO GdRG set. The other two sets of fragmentation functions
include higher order corrections beyond the formal accuracy of the calculation.  Although including $\mathcal{O}(\alpha\alpha_s)$ 
fragmentation functions captures part of the NNLO corrections to triphoton production, other contributions -- such as those associated
with two LO fragmentation processes -- are not included.  This fact may explain the unusual behaviour of the predictions for $R_0=0.4$,
where for $\epsilon_{\gamma} < 0.5$ the cross section decreases as $\epsilon_{\gamma}$ increases. 
The decrease in cross section is much more pronounced for the $\mathcal{O}(\alpha\alpha_s$) set of GdRG.
It is tempting to conclude from Fig.~\ref{fig:epscomp} that higher order corrections could be
sizeable, but a priori we do not know the effect of the remaining higher order contributions. Therefore we advocate the use of the $\mathcal{O}(\alpha)$
fragmentation  functions for NLO predictions at the LHC.  In this case we observe that such predictions are close to those
obtained using smooth cone isolation.  This suggests that, for cuts that are similar to the ones we have used here, 
the use of smooth cone isolation for this process should provide an adequate description of experimental isolation requirements.

Comparing the results for different cone sizes it is clear that the cross sections obtained using the larger cone size $R_0=0.7$ depend
much more strongly on $\epsilon_{\gamma}$. This reflects the importance of the real radiation terms on the total cross section. For 
large values of $\epsilon_{\gamma}$ the cross section obtained using smooth cone isolation is more strongly suppressed than 
for the fractional isolation. 
This suppression can be explained by considering event topologies in which a radiated parton 
is near the threshold for acceptance in the inner cone. 
In these topologies we assume that the radiation in the smaller cone (for example $R_0=0.4$) is just soft enough  
to pass the isolation requirement. For the fractional isolation this event will then pass all subsequent 
increases in cone size, since the parameters used to determine the isolation requirements are fixed (the 
transverse momenta of the parton and of the photon). However, for the smooth cone isolation the isolation requirements 
for this event change as a function of the cone size, due to the $(1-\cos{R_0})$ pre-factor in Eq.~(\ref{eq:smooth}).
Therefore as the cone size increases the smooth cone isolation requirement becomes tighter and thus more events are
rejected than in the fractional isolation case. 

\begin{center}
\begin{figure} 
\includegraphics[width=8cm]{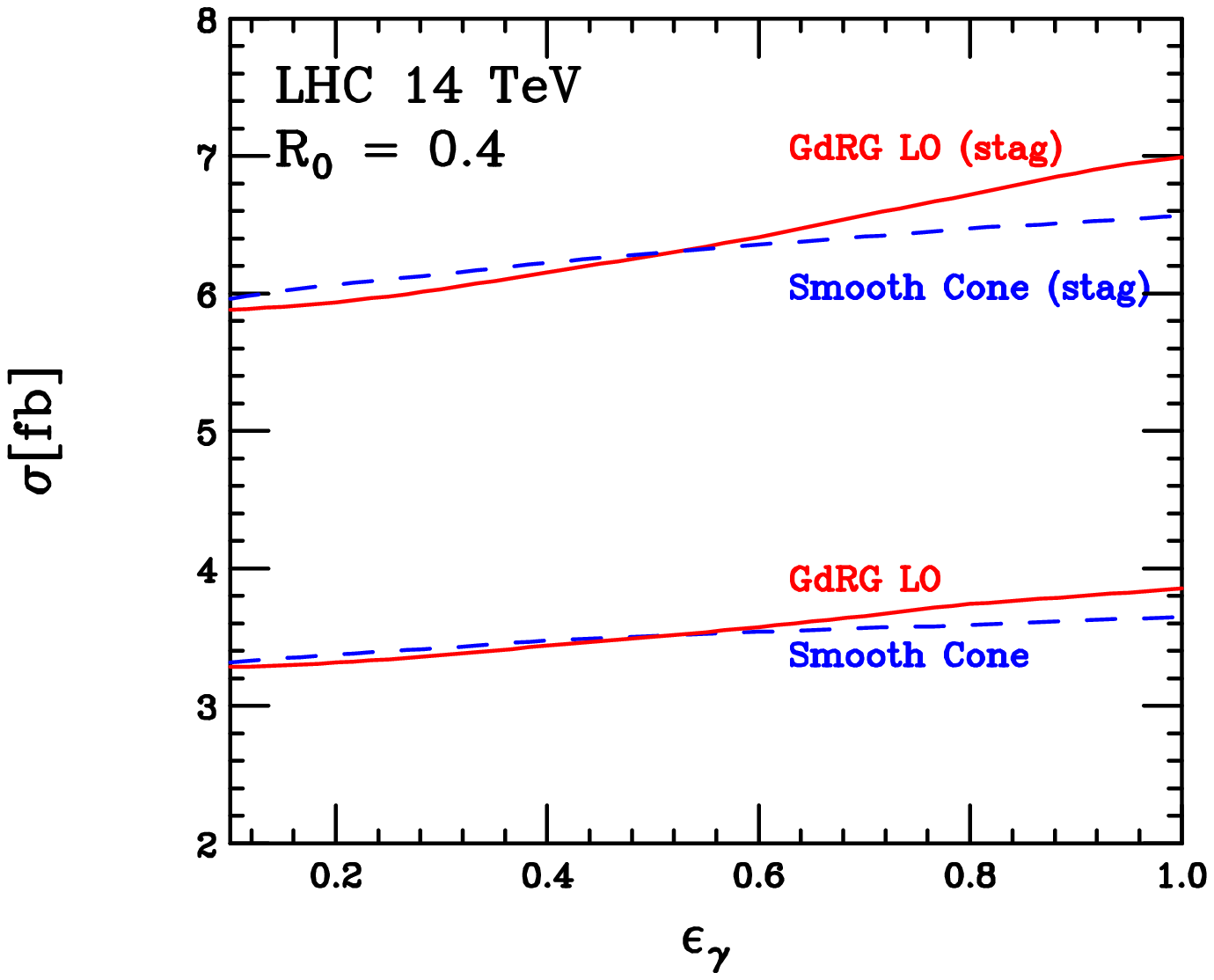}
\includegraphics[width=8cm]{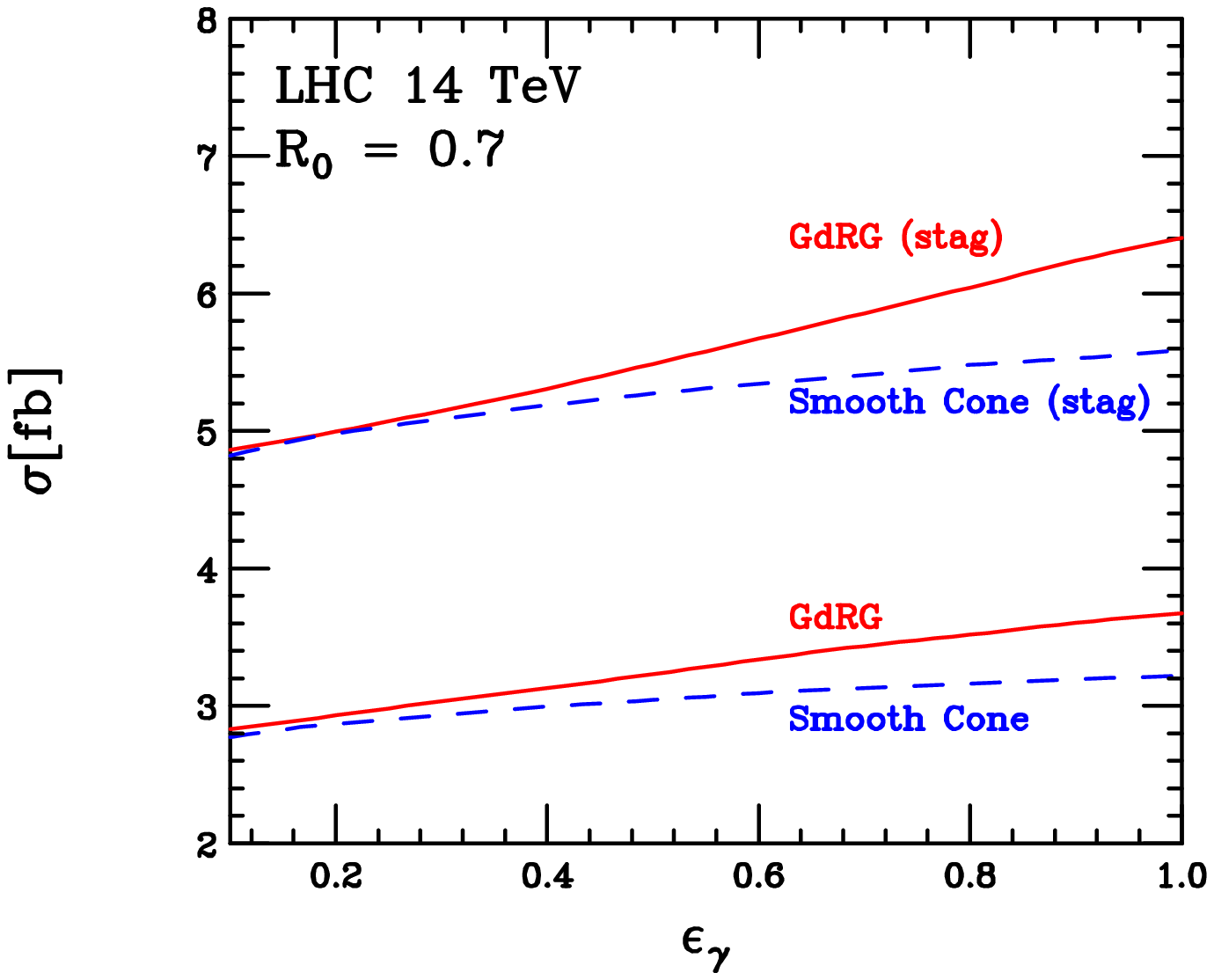}
\caption{Dependence of the NLO triphoton cross section on the parameter that controls the amount of hadronic energy inside
the isolation cone, $\epsilon_\gamma$ with harder selection requirements $p_T^{\gamma} > 50$ GeV, and staggered cuts $p_T^{\gamma} > 100, 50, 30$ GeV.  Results are shown for the fractional and smooth cone isolation procedures, using
an isolation cone of size $R_0=0.4$ (left) and $R_0=0.7$ (right).}
\label{fig:epscomp_50}
\end{figure} 
\end{center}
In order to check the dependence of the isolation algorithms on the event selection cuts. we have repeated this analysis 
using selection criteria with higher cuts on the photon transverse momenta. 
Specifically, we modify the photon transverse momentum cuts of Eq.~(\ref{eq:basiccuts}) as follows, with the other cuts unchanged.
In the first case we simply raise the cut uniformly and require $p_T^{\gamma} > 50$~GeV.  In the second case we use a set of cuts with
staggered thresholds, $p_T^{\gamma_1} > 100$~GeV, $p_T^{\gamma_2} > 50$~GeV and $p_T^{\gamma_3} > 30$~GeV where the photons
are labelled according to $p_T^{\gamma_1} > p_T^{\gamma_2} > p_T^{\gamma_3}$. Our results are shown in Fig.~\ref{fig:epscomp_50}. It is clear from comparing
Figs.~\ref{fig:epscomp} and~\ref{fig:epscomp_50} that the overall structure of the results remains the same.  The smooth cone algorithm
is in reasonable agreement with the fractional isolation result using the LO GdRG fragmentation functions. There is particularly good agreement 
for the smaller cone size of $R_0=0.4$, whilst more significant differences are observed for the larger cone choice and $\epsilon_{\gamma} > 0.5$. 
As the photon transverse momentum cut is raised, the smooth cone results depend less strongly on $\epsilon_{\gamma}$ than those including 
fragmentation. This is illustrated by the fact that, for $R_0=0.4$, the GdRG prediction is smaller than the smooth cone result
for $\epsilon_{\gamma} < 0.5$ but higher for $\epsilon_{\gamma} > 0.5$. The exact value of $\epsilon_{\gamma}$ for which 
the predictions intersect is of course dependent on the phase space selection requirements. For example, for the loose cuts defined previously 
the predictions intersected around $\epsilon_{\gamma} = 0.95$ for $R_0=0.4$, c.f. Fig.~\ref{fig:epscomp}.

\renewcommand{\baselinestretch}{1.2}
\begin{table}
\begin{tabular}{|c|l||ccc|ccc|}
\hline
 &  & \multicolumn{3}{c|}{\parbox[c]{0.3\linewidth}{$R_0=0.4$}} & 
\multicolumn{3}{c|}{\parbox[c]{0.3\linewidth}{$R_0=0.7$}} \\
min. $p_T^\gamma$
& \qquad \quad isolation  & \parbox[c]{0.1\linewidth}{$E=5$} 
                          & \parbox[c]{0.1\linewidth}{$E=25$} 
                          & \parbox[c]{0.1\linewidth}{$E=50$}
                          & \parbox[c]{0.1\linewidth}{$E=5$}
                          & \parbox[c]{0.1\linewidth}{$E=25$}
                          & \parbox[c]{0.1\linewidth}{$E=50$} \\
\hline
\hline
$30$ GeV &
 fixed, $E_T^{\rm max}=E$ [GeV]      & 16.86 & 17.56 & 19.45 & 14.16 & 16.00 & 18.61 \\
& fractional, $\epsilon_\gamma=E/30$ & 16.96 & 18.76 & 21.15 & 14.43 & 17.48 & 20.51 \\
& smooth., $\epsilon_\gamma=E/30$    & 17.58 & 19.00 & 20.15 & 14.58 & 16.37 & 17.60 \\
 \hline
$50$ GeV &
 fixed, $E_T^{\rm max}=E$ [GeV]      & 3.26  & 3.37 & 3.60  & 2.76 & 3.04 & 3.39 \\
& fractional, $\epsilon_\gamma=E/50$ & 3.28  & 3.50 & 3.86  & 2.83 & 3.23 & 3.68 \\
& smooth., $\epsilon_\gamma=E/50$    &  3.32 & 3.51 & 3.65  & 2.77 & 3.04 & 3.22 \\
 \hline
\end{tabular}
\renewcommand{\baselinestretch}{1}
\caption{Triphoton cross sections at the LHC (in femtobarns), computed using the
fixed energy, fractional energy and smooth cone forms of isolation prescription.
The comparison uses the LO GdRG fragmentation functions and is performed for two
values of the photon $p_T$ cut.
\label{tab:fixE}}
\end{table}

Finally we turn to the case of fixed energy isolation.
In Table~\ref{tab:fixE} we present results obtained using this form of isolation and
compare them to the cross sections obtained using fractional and smooth cone isolation.  Specifically,
for fixed isolation with a maximum transverse energy $E_T^{\rm max}$ in Eq.~(\ref{eq:fixedE}),
we compare to fractional and smooth cone isolation with $\epsilon_{\gamma} = E_T^{\rm{max}} / p_{T,min}^{\gamma}$
in Eq.~(\ref{eq:fracE}). When the isolation is tight ($E=5$ in Table~\ref{tab:fixE}),
 the results obtained in the different cases are in very good agreement,
which is simply a reflection of the fact that most of the cross section is due to production of photons near the minimum $p_T$ threshold.  
However, as the isolation requirement weakens, the predictions begin to show bigger differences. Requiring a much looser
criterion, $E=50$, induces differences
of up to 10\% for fractional and fixed isolation, with slightly smaller differences between smooth cone and fixed isolation.

\subsection{Isolation effects in $\gamma\gamma+$jet production}

\begin{center}
\begin{figure} 
\includegraphics[width=8cm]{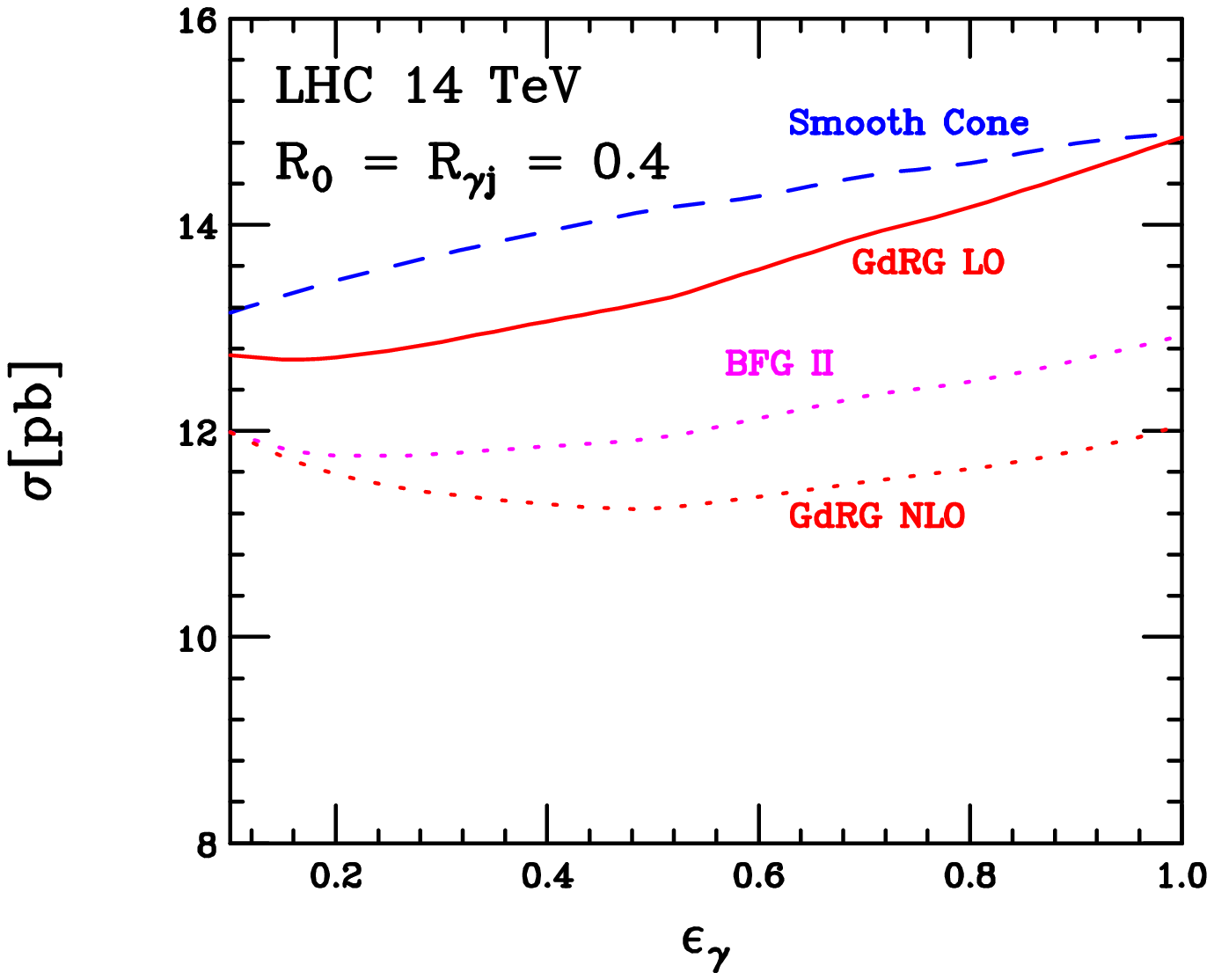}
\includegraphics[width=8cm]{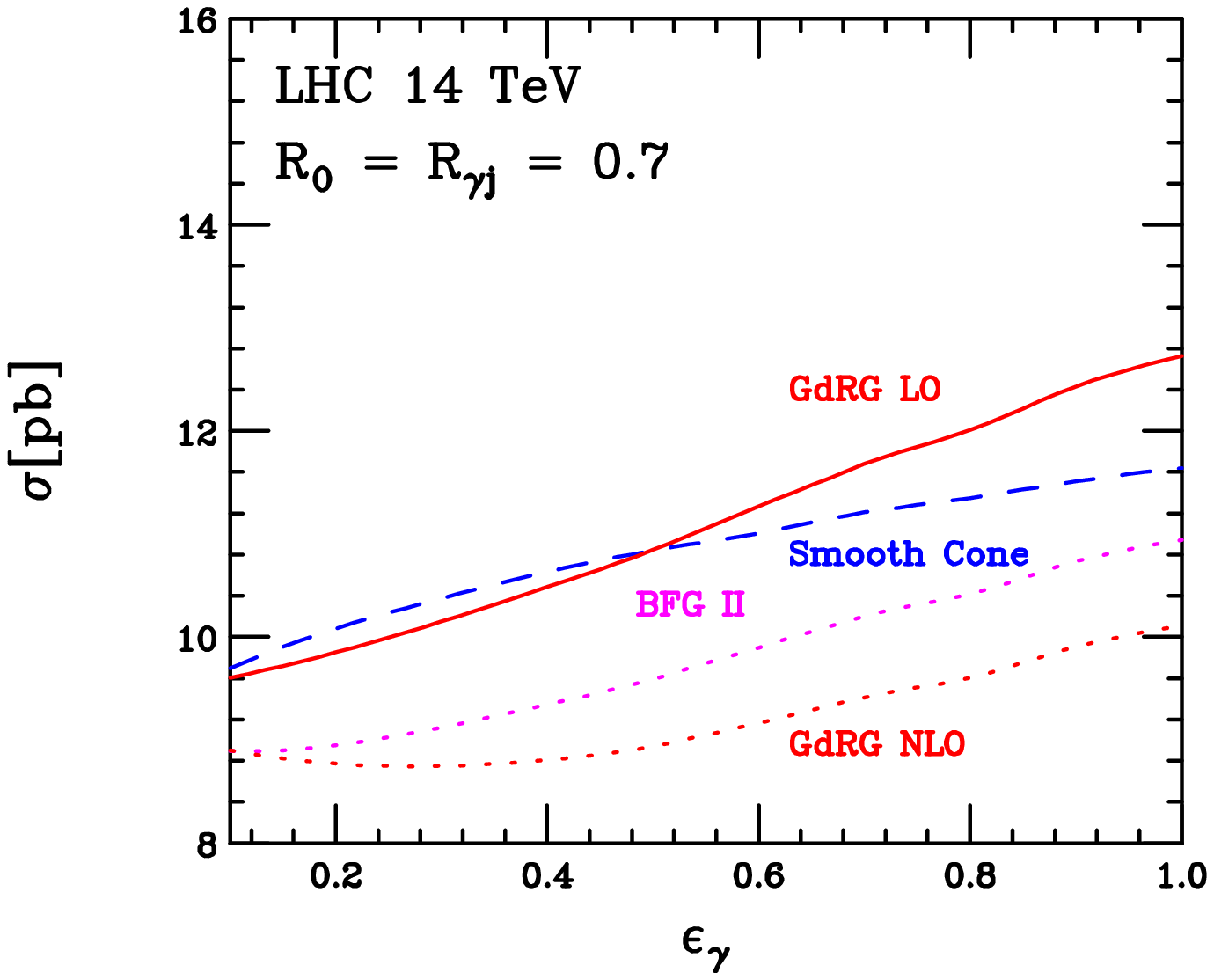}
\caption{Dependence of the NLO $\gamma\gamma+$jet cross section on the parameter that controls the amount of hadronic energy inside
the isolation cone, $\epsilon_\gamma$.  Results are shown for the fractional and smooth cone isolation procedures, using
an isolation cone of size $R_0=0.4$ (left) and $R_0=0.7$ (right).}
\label{fig:epscomp_gagajet}
\end{figure} 
\end{center}
As already noted, it is interesting to compare the isolation effects in $\gamma\gamma\gamma$ and $\gamma\gamma+$jet processes.
In order to maximize the similarities with the triphoton results that have just been presented, for the $\gamma\gamma+$jet final state 
we adopt the same photon cuts as in Eq.~(\ref{eq:basiccuts}) and tailor the jet cuts as follows.  Partons are clustered into jets using the
anti-$k_T$ algorithm with $D=0.5$ and are required to satisfy the same rapidity requirement as the photons, i.e. $|\eta_{j}| < 2.5$.
To obtain a finite cross section we must impose a minimum jet-photon separation, $R_{\gamma,j}$.  For this we use the same value as
for the isolation cone, namely  $R_{\gamma,j} = R_0 = \{0.4, 0.7\}$.  The common scale choice, $\mu$, that we have used for these calculations
is given by $\mu^2=m_{\gamma\gamma}^2 + \sum (p^j_{T})^2$.

Results for the NLO cross section as a function of $\epsilon_{\gamma}$ for $p^j_T , p_T^{\gamma} > $ 30 GeV are shown
in Fig.~\ref{fig:epscomp_gagajet}, for the two choices $R_0=0.4$ and $R_0=0.7$.
The predictions for $\gamma\gamma+$jet production are very similar to the equivalent results obtained
for the $\gamma\gamma\gamma$ process (c.f. Fig.~\ref{fig:epscomp}), suggesting that the 
dependence of the cross section on $\epsilon_{\gamma}$ is not strongly influenced by the number of photons.
Instead it is governed by the kinematics of the underlying scattering. 

We note that a similar study was undertaken in ref.~\cite{Gehrmann:2013aga} employing selection cuts relevant for 
Higgs searches in the channel $H \to \gamma\gamma$.
We have repeated this analysis using the results of this paper and find agreement for very tight
isolation requirements but substantial differences for larger values of $\epsilon_{\gamma}$.  The
qualitative behaviour
of our predictions is more similar to the results shown in Fig.~\ref{fig:epscomp_gagajet},
with a milder dependence of the cross section on $\epsilon_\gamma$.
We understand that this difference is due to an error
in the implementation of the fragmentation functions in ref.~\cite{Gehrmann:2013aga}.
\footnote{We thank the authors of ref.~\cite{Gehrmann:2013aga} for confirmation of this issue.}

\subsection{Summary}

It is clear from the results of this section that the predictions using smooth cone isolation and those using
fractional isolation are in reasonable agreement with one another, provided that the fragmentation functions are
restricted to fixed $\mathcal{O}(\alpha)$ accuracy. 
The agreement is particularly good for smaller cone choices and tighter isolation requirements. 
For smooth cone isolation with larger cones the $(1-\cos{R_0})^{-1}$ prefactor tightens the isolation, and 
results in larger differences between smooth cone and fractional energy isolation for the same choice of 
$\epsilon_{\gamma}$.  For all of the phase space selection cuts 
we investigated, the smooth cone results showed the mildest dependence on $\epsilon_{\gamma}$.
Therefore, varying $\epsilon_\gamma$ in a smooth cone calculation in order to gauge the uncertainty
associated with isolation effects is not advisable.
We observed that including higher order corrections to the fragmentation functions 
induced large differences with respect to the smooth cone and LO GdRG sets. This may be indicative of 
large NNLO corrections, but since they are only a partial computation no definitive statement can be made 
on the impact of higher order corrections.  

\section{Results} 
\label{sec:results}

\subsection{Triphotons at the LHC} 
\label{sec:LHC}

In this section we provide predictions for the triphoton process at the LHC, operating
at a variety of center of mass energies.  We use a set of basic cuts that is appropriate for
experimental analyses at the LHC and which closely corresponds to the cuts used in the previous
section, c.f. Eq.~(\ref{eq:basiccuts}).  The photons are required to satisfy,
\begin{eqnarray} 
|\eta_{\gamma}| < 2.5 \;, \qquad R_{\gamma\gamma} > 0.4 \;, \qquad
\sum_{\in R_{\gamma}=0.4} E_T^{\rm{had}}  <  0.4 \;p_{T}^{\gamma} \;,
\end{eqnarray}
i.e. we use the fractional form of isolation and, following the conclusions of our previous analysis,
the LO GdRG set of fragmentation functions.  As before we 
employ the CT10 (CTEQ6L1) PDF set for our NLO (LO) predictions.
We consider two thresholds for the photon transverse momenta, $p_T^{\gamma} > 30$~GeV and
$p_T^{\gamma} > 50$~GeV.  Our results for the two values of the cut are
collected in Table~\ref{tab:xs_LHC}.
\renewcommand{\baselinestretch}{1.4}
\begin{table}
\begin{tabular}{|cc||c|c|c|}
\hline
\qquad $\sqrt{s}$ \qquad \quad & photon cut & \parbox[c]{0.12\linewidth}{LO [fb]} & \parbox[c]{0.12\linewidth}{NLO [fb]}
 & \parbox[c]{0.12\linewidth}{$K$-factor} \\
\hline
\hline
7 TeV  & $p_T^{\gamma} > 30$ GeV & $3.36_{-2\%}^{+1\%}$  & 7.49$^{+6\%}_{-4\%}$  & 2.23  \\
 & $p_T^{\gamma} > 50$ GeV & $0.64_{-1\%}^{+2\%} $ & 1.30$^{+6\%}_{-5\%}$  & 2.03 \\
\hline
8 TeV  & $p_T^{\gamma} > 30$ GeV & $3.89_{-3\%}^{+2\%} $ & 8.87$^{+5\%}_{-5\%}$ & 2.28  \\
 & $p_T^{\gamma} > 50$ GeV & $ 0.77_{-1\%}^{+1\%}  $       & 1.60$^{+6\%}_{-4\%}$  &2.08  \\
\hline
13 TeV  & $p_T^{\gamma} > 30$ GeV & $6.42_{-5\%}^{+4\%} $ &15.87$^{+4\%}_{-3\%}$  & 2.47 \\
 & $p_T^{\gamma} > 50$ GeV &   $ 1.38_{-1\%}^{+1\%}$       & 3.13$^{+5\%}_{-4\%}$  & 2.27  \\
\hline
14 TeV  & $p_T^{\gamma} > 30$ GeV & $6.91_{-6\%}^{+5\%} $ & 17.28$^{+4\%}_{-3\%}$ & 2.50 \\
 & $p_T^{\gamma} > 50$ GeV &  $1.50_{-2\%}^{+1\%}$ &  3.44$^{+5\%}_{-4\%}$& 2.29 \\
\hline
\end{tabular}
\renewcommand{\baselinestretch}{1}
\caption{Summary of LHC triphoton cross sections at various LHC operating energies, with
two choices of photon $p_T$ threshold. 
\label{tab:xs_LHC}}
\end{table}

The results have been obtained using $m_{\gamma\gamma\gamma}$ as the central 
renormalization, factorization and fragmentation scale and the quoted uncertainty corresponds to variation of this 
central scale by a factor of two in each direction.  Since this process does not depend on the strong coupling
at leading order, there is only a very small dependence on the factorization scale at that order.
At NLO the prediction becomes sensitive to the gluon distribution and, as a result,
we observe large $K$-factors ($\sim 2-2.5)$ when going from 
LO to NLO.  Thus it is only at NLO that one obtains a realistic prediction
for the normalization of these processes at the LHC.
At NLO the scale dependence remains rather small, and reflects a partial cancellation between the factorization 
and renormalization scales. As $\sqrt{s}$ increases the dependence on the factorization scale increases,
as can clearly be seen from the LO results, such that the cancellation becomes more complete.
At 14 TeV this procedure yields a scale uncertainty of about $\pm 4\%$.
It should be borne in mind that other sources of uncertainty, for instance due to the particular choice of
fragmentation functions, are not accounted for here.  As noted in the previous section such uncertainties
may be at least as large.

\begin{center}
\begin{figure} 
\includegraphics[width=10cm]{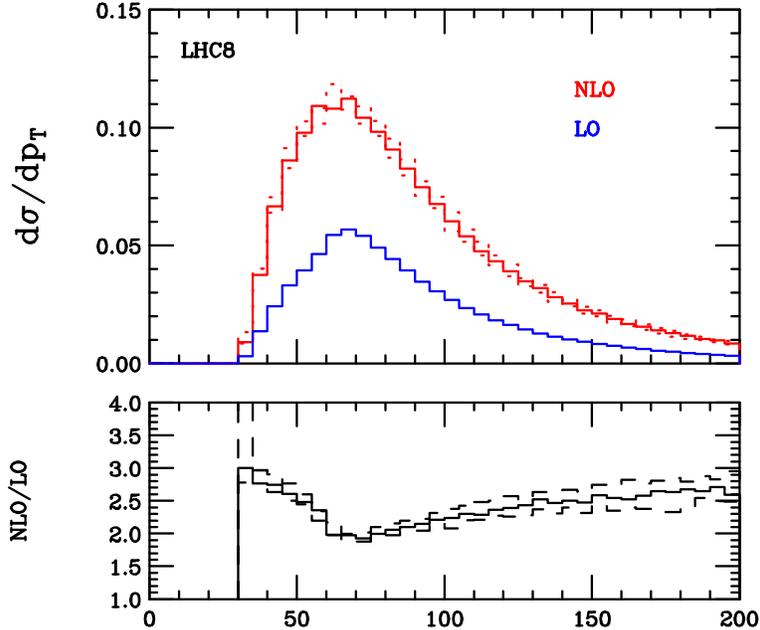}
\caption{The $p_{T,\gamma}$ spectrum for the hardest photon at the 8 TeV LHC. The solid lines
represent the contributions with $\mu=m_{\gamma\gamma\gamma}$, the dashed lines represent the 
NLO predictions with $\mu=\{0.5,2\} m_{\gamma\gamma\gamma}$.} 
\label{fig:LHCPT}
\end{figure} 
\end{center}

In Fig.~\ref{fig:LHCPT} we present the differential distribution for the $p_T$  of the hardest photon.
This distribution is significantly 
altered by the higher order corrections, both in rate and shape. The region $p_T < 2 p_T^{min}$ experiences
the most dramatic corrections. Since we require three photons with $p_T > 30$~GeV, this distribution has a distinct
broad peak around $60$~GeV.  At NLO the kinematic suppression in the region $p_T<60$~GeV is reduced
due to the presence of real radiation contributions that allow a parton to recoil against the photonic system.
This leads to the $K$-factor in this region being larger than at the peak of the distribution. 

\subsection{Triphotons at the Tevatron}
\label{sec:Tev}
Since the leading order triphoton process is $q{\bar q} \to \gamma\gamma\gamma$, this process is 
significantly easier to produce at a $p\overline{p}$ collider where both the quark and the antiquark may be described
by the valence content of the proton and antiproton respectively.  This is to be contrasted with
an equivalent $pp$ machine where the antiquark must be obtained from the proton sea. For this 
reason it is interesting to contrast the LHC results of the previous section with the triphoton
cross section at the Tevatron.

To that end we define a set of cuts that is appropriate for experimental analyses that could be
performed at the Tevatron, 
\begin{eqnarray} 
p^{\gamma}_{T} > 15 \; {\rm{GeV}}\;, \qquad |\eta_{\gamma}| < 1.1 \;, \qquad R_{\gamma\gamma} > 0.4 \;,\qquad
\sum_{\in R_{\gamma}=0.4} E_T^{\rm{had}}  < 5 \; {\rm{GeV}}
\end{eqnarray}
and use the same parton distribution and fragmentation functions as before.  Note though that the
isolation condition is now at a fixed energy rather than taking the fractional form, although since
the isolation condition is rather strict it could be well-described by a corresponding fractional
isolation, c.f. Table~\ref{tab:fixE}.
With these cuts the triphoton cross section is, 
\begin{eqnarray} 
\sigma_{\gamma\gamma\gamma}^{NLO} = 4.74_{-5\%}^{+6\%} \;  \rm{[fb]}
\label{eq:sigbasTev}
\end{eqnarray} 
As was the case at the LHC, the NLO corrections to this process result in a large $K$-factor (1.93) when 
comparing with the LO result of $2.46$fb (obtained using the CTEQ6L1 PDF set).
The upper and lower percentages in the above 
result represent scale variation by a factor of two around a central scale 
choice of $\mu=m_{\gamma\gamma\gamma}$. 
Given the 10 fb$^{-1}$ of data recorded by the CDF and D0 detectors during Run II of the Tevatron,
one therefore expects $\mathcal{O}(50)$ events in the total data set, before accounting for
acceptance and efficiency corrections. This suggests that 
a measurement of this process by the Tevatron experiments may therefore
also be possible.

\section{Conclusions}
\label{sec:conc}

In this paper we have studied the production of triphoton final states at hadron colliders, paying 
particular attention to the role of photon fragmentation and isolation effects. We calculated compact
analytic expressions for $\gamma\gamma\gamma$ and $\gamma\gamma+$jet one-loop amplitudes and used them
to implement these processes in MCFM. We investigated the impact
of different isolation prescriptions, finding that 
the NLO cross section is quite sensitive to the type of isolation and fragmentation functions applied. 
This is due to the fact that a large part of the NLO cross section arises from configurations that contain 
an initial state gluon and, consequently, a quark in the final state.  With three photons in the 
final state, the phase space in which this quark is near a photon is large and therefore the NLO prediction is 
especially sensitive to the isolation definition. We investigated the impact of different fragmentation functions on the NLO 
cross section. We found that the results obtained using predictions accurate to $\mathcal{O}(\alpha^3\alpha_s)$, i.e. using
smooth cone isolation and LO GdRG fragmentation contributions, are similar to each other. However, including higher order effects in the predictions
for the fragmentation functions resulted in substantially different results.
Since our predictions do not include the full $\mathcal{O}(\alpha_s^2)$
corrections we advocated the use of LO fragmentation sets which result in a consistent NLO prediction.

We studied the phenomenology of triphoton production at hadron colliders. We presented NLO cross sections for 
a range of LHC operating energies and phase space selection criteria. At 8 TeV the cross sections 
are typically around $5$~fb, suggesting that this signal should be observable in the existing LHC data set. The measurement 
of this cross section would be one of the first studying triboson production.  The study of this class of processes is 
extremely interesting since it tests the interplay between electroweak and QCD physics in a final
state involving multiple electroweak couplings. 

Since at LO the production of triphotons is dominated by $u\overline{u}$ initial states, $p\overline{p}$ colliders 
are more sensitive to this process than equivalent $pp$ machines. Indeed the cross section at the Tevatron,
with different cuts more appropriate for a lower-energy machine, is also around $5$~fb.
Therefore with the 10 fb$^{-1}$ data set accumulated by 
the CDF and D0 experiments this signature may also be observable at the Tevatron.

Our results for $\gamma\gamma\gamma$ and $\gamma\gamma+$jet have been implemented into MCFM v6.8, including 
all of the fragmentation functions studied in this paper. 

\noindent
\section*{Acknowledgements}
We thank Keith Ellis and the authors of ref.~\cite{Gehrmann:2013aga} for useful discussions.
The research of J.M.C. is supported by the U.S. DOE under contract DE-AC02-07CH11359.

\appendix

\section{Amplitudes for $\gamma\gamma$+jet}

In this appendix we present the amplitudes needed to compute the NLO corrections to
$\gamma\gamma+$jet production,
\begin{equation}
0 \longrightarrow {\bar q}(p_1) + q(p_2) + g(p_3) +\gamma(p_4)+ \gamma(p_5) \;.
\end{equation}
At leading order the amplitude can be decomposed into the following color stripped amplitude, 
\begin{eqnarray} 
A^{(0)}(1_{\overline{q}}^{h_1},2_q^{h_2},3_{g}^{h_3},4_{\gamma}^{h_4},5_{\gamma}^{h_5})= 2\sqrt{2} i \;Q_q^2 e^2 g_s (T^{a_3}_{i_1i_2})
\mathcal{A}^{(0)}(1_{\overline{q}}^{h_1},2_q^{h_2},3_{g}^{h_3},4_{\gamma}^{h_4},5_{\gamma}^{h_5})
\end{eqnarray}
The non-vanishing helicity amplitudes for the LO process have identical kinematic structure to Eq.~(\ref{eq:lohel}), i.e. 
\begin{eqnarray} 
\mathcal{A}^{(0)}(1_{\overline{q}}^-,2_q^+,3_{g}^+,4_{\gamma}^+,5_{\gamma}^-)
 = \frac{\spa1.2\spa1.5^2}{\spa1.3\spa1.4\spa2.3\spa2.4} \;, \\
\mathcal{A}^{(0)}(1_{\overline{q}}^-,2_q^+,3_{g}^-,4_{\gamma}^+,5_{\gamma}^+)
 = \frac{\spa1.2\spa1.3^2}{\spa1.5\spa1.4\spa2.5\spa2.4} \;.
\end{eqnarray}
For simplicity we refer to these helicity structures as the ``$\gamma$-MHV" and ``$g$-MHV" amplitudes, with the
nomenclature denoting the identity of the negative-helicity particle.
Although these amplitudes are trivially related at LO,
\begin{equation}
\mathcal{A}^{(0)}(1_{\overline{q}}^-,2_q^+,3_{g}^-,4_{\gamma}^+,5_{\gamma}^+)
 = \mathcal{A}^{(0)}(1_{\overline{q}}^-,2_q^+,5_{g}^+,4_{\gamma}^+,3_{\gamma}^-) \;,
\end{equation}
differences arise in the one-loop and real radiation amplitudes. 
At one loop the amplitude can be decomposed into the following leading and subleading color pieces 
\begin{eqnarray}  
&& A^{(1)}(1_{\overline{q}}^{h_1},2_q^{h_2},3_{g}^{h_3},4_{\gamma}^{h_4},5_{\gamma}^{h_5}) =N_c \frac{\alpha_s}{4\pi}(2\sqrt{2}) i \; e^2 g_s(T^{a_3}_{i_1i_2})
\bigg(Q_q^2\mathcal{A}^{(L)}(1_{\overline{q}}^{h_1},2_q^{h_2},3_{g}^{h_3},4_{\gamma}^{h_4},5_{\gamma}^{h_5})\nonumber\\&&+\frac{Q_q^2}{N_c^2}
\mathcal{A}^{(R)}(1_{\overline{q}}^{h_1},2_q^{h_2},3_{g}^{h_3},4_{\gamma}^{h_4},5_{\gamma}^{h_5}) 
-\sum_{i}\frac{ Q_i^2}{N_c}\mathcal{A}^{N_f}(1_{\overline{q}}^{h_1},2_q^{h_2},3_{g}^{h_3},4_{\gamma}^{h_4},5_{\gamma}^{h_5})\bigg)
\end{eqnarray}
In the above equation the summation over $i$ represents the contributions arising from closed (light) fermion loops, which in our implementation corresponds to $i=u,d,s,c,b$. In all cases the subleading $R$ pieces can be obtained from the $\gamma\gamma\gamma$ helicity amplitudes presented in section~\ref{sec:calc}, i.e. $
\mathcal{A}^{(R)}(1_{\overline{q}}^{h_1},2_q^{h_2},3_{g}^{h_3},4_{\gamma}^{h_4},5_{\gamma}^{h_5}) = \mathcal{A}^{(1)}(1_{\overline{q}}^{h_1},2_q^{h_2},3_{\gamma}^{h_3},4_{\gamma}^{h_4},5_{\gamma}^{h_5}) $.
As a result we need only present the amplitudes that appear at leading color or contain closed fermion loops.

We begin by presenting the unrenormalized $\gamma$-MHV amplitude,
\begin{eqnarray}
&&\mathcal{A}^{(L)}(1_{q}^-,2_{\overline{q}}^+,3_g^+,4_{\gamma}^+,5_{\gamma}^-)=
\nonumber\\&&\bigg[\frac{1}{\epsilon^2}\bigg(\bigg(\frac{\mu^2}{-s_{13}}\bigg)^{\epsilon}
+\bigg(\frac{\mu^2}{-s_{23}}\bigg)^{\epsilon}\bigg)+\frac{3}{2\epsilon}\bigg(\frac{\mu^2}{-s_{25}}\bigg)^{\epsilon} + 3 \bigg]
\mathcal{A}^{(0)} (1_{q}^-,2_{\overline{q}}^+,3_g^+,4_{\gamma}^+,5_{\gamma}^-)\nonumber\\
&&-\frac{\spa1.2^3\spa4.5^2}{\spa1.3\spa1.4\spa2.3\spa2.4^3}\Ls_{-1}\bigg(s_{13};s_{45},s_{25}\bigg)
  +\frac{\spa1.5^2}{\spa1.3\spa2.4\spa3.4}\Ls_{-1}\bigg(s_{15};s_{24},s_{23}\bigg)\nonumber\\
&&-\frac{\spa1.2\spa1.5^2}{\spa1.3\spa1.4\spa2.3\spa2.4}\Ls_{-1}\bigg(s_{23};s_{45},s_{15}\bigg)
  -\frac{\spa1.3^2\spa4.5^2}{\spa1.4\spa2.3\spa3.4^3}\Ls_{-1}\bigg(s_{25};s_{14},s_{13}\bigg)\nonumber\\
&&-\frac{\spa1.2\spa1.5^2}{\spa1.3\spa1.4\spa2.3\spa2.4}\Ls_{-1}\bigg(s_{45};s_{13},s_{23}\bigg)
  +\frac{\spa1.2\spa2.5^2\spb3.2}{\spa2.3\spa2.4^2}\frac{L_0(-s_{13},-s_{45})}{s_{45}}\nonumber\\
&&-\frac{(\spa1.3\spa2.4+\spa1.2\spa3.4)\spa4.5^2\spb4.3}{\spa2.4^2\spa3.4^2}\frac{L_0(-s_{13},-s_{25})}{s_{25}}
  -\frac{\spa1.3\spa4.5^2\spb4.3^2}{2\spa2.4\spa3.4}\frac{L_1(-s_{13},-s_{25})}{s_{25}^2}\nonumber\\
&&+\left(\spa1.5\spa3.4-\spa1.3\spa4.5\right)
   \frac{\spa3.5\spb4.3}{\spa2.3\spa3.4^2}\frac{L_0(-s_{14},-s_{25})}{s_{25}}
  -\frac{\spa1.4\spa3.5^2\spb4.3^2}{2\spa2.3\spa3.4}\frac{L_1(-s_{14},-s_{25})}{s_{25}^2}\nonumber\\
&&+\frac{\spa1.2\spa2.5\spa1.5}{\spa1.3\spa2.3\spa2.4^2}\log\bigg({\frac{s_{45}}{s_{25}}\bigg)}
  +\frac{\spa3.4\spb3.4^2}{2\spa2.3\spa2.4\spb1.5\spb2.5} \;.
\end{eqnarray}
The unrenormalized $g$-MHV amplitude is given by, 
\begin{eqnarray}
&&\mathcal{A}^{(L)}(1_{q}^-,2_{\overline{q}}^+,3_g^-,4_{\gamma}^+,5_{\gamma}^+)=\nonumber\\&&
\bigg[\frac{1}{\epsilon^2}\bigg(\bigg(\frac{\mu^2}{-s_{13}}\bigg)^{\epsilon}
+\bigg(\frac{\mu^2}{-s_{23}}\bigg)^{\epsilon}\bigg)+\frac{3}{2\epsilon}\bigg(\frac{\mu^2}{-s_{23}}\bigg)^{\epsilon} + 3 \bigg]
\mathcal{A}^{(0)}(1_{q}^-,2_{\overline{q}}^+,3_g^-,4_{\gamma}^+,5_{\gamma}^+)\nonumber\\
&&+\frac{\spa1.3^2}{\spa1.4\spa2.5\spa4.5}\Ls_{-1}\bigg(s_{14};s_{25},s_{23}\bigg)
  -\frac{\spa1.3^2}{\spa1.5\spa2.4\spa4.5}\Ls_{-1}\bigg(s_{15};s_{24},s_{23}\bigg) \nonumber\\
&&-\frac{\spa1.3^2}{\spa1.5\spa2.4\spa4.5}\Ls_{-1}\bigg(s_{24};s_{15},s_{13}\bigg)
  +\frac{\spa1.3^2}{\spa1.4\spa2.5\spa4.5}\Ls_{-1}\bigg(s_{25};s_{14},s_{13}\bigg)\nonumber\\
&&-\frac{\spa1.4\spa3.5^2\spb5.4^2}{2\spa2.5\spa4.5}\frac{L_1(s_{14},s_{23})}{s_{23}^2}
  -\frac{\spa1.3\spa3.5\spb5.4}{\spa2.5\spa4.5}\frac{L_0(s_{14},s_{23})}{s_{23}}\nonumber\\
&&+\frac{\spa1.5\spa3.4^2\spb5.4^2}{2\spa2.4\spa4.5}\frac{L_1(s_{15},s_{23})}{s_{23}^2}
  -\frac{\spa1.3\spa3.4\spb5.4}{\spa2.4\spa4.5}\frac{L_0(s_{15},s_{23})}{s_{23}}\nonumber\\
&&-\frac{\spb4.5}{2\spb1.3\spb2.3}\left(\frac{\spb2.4}{\spa2.5}-\frac{\spb2.5}{\spa2.4}\right)
  +\frac{\spa1.3\spb4.5}{2\spa2.3\spb2.3\spa4.5}\left(\frac{\spa3.4}{\spa2.4}+\frac{\spa3.5}{\spa2.5}\right) \;.
\end{eqnarray}
These amplitudes must be renormalized, which is achieved 
by adding a UV counterterm that also accounts for the transition to the normal $\overline{\rm{MS}}$ definition
of the strong coupling, 
\begin{eqnarray} 
&&\mathcal{A}^{(L)}(1_{\overline{q}}^{h_1},2_q^{h_2},3_{g}^{h_3},4_{\gamma}^{h_4},5_{\gamma}^{h_5})  \to \nonumber \\
&& \quad \mathcal{A}^{(L)}(1_{\overline{q}}^{h_1},2_q^{h_2},3_{g}^{h_3},4_{\gamma}^{h_4},5_{\gamma}^{h_5}) 
-\frac{1}{6}\left[\frac{1}{\epsilon}\left(11-\frac{2N_f}{N_c}\right)-1\right]
 \mathcal{A}^{(0)}(1_{\overline{q}}^{h_1},2_q^{h_2},3_{g}^{h_3},4_{\gamma}^{h_4},5_{\gamma}^{h_5}) \;.
\end{eqnarray}

Finally the amplitude arising from closed loops of fermions is,
\begin{eqnarray}
&&\mathcal{A}^{N_f}(1_{q}^-,2_{\overline{q}}^+,3_g^+,4_{\gamma}^+,5_{\gamma}^-)
=-2\left(\frac{\spa1.4^2\spa3.5^2+\spa1.3^2\spa4.5^2}{\spa1.2\spa3.4^4}\right)
   \Ls_{-1}(s_{12}; s_{35},s_{45}) \nonumber\\
&&-\frac{\spa1.4\spa3.5\spb4.3}{\spa1.2\spa3.4^3} \left(2\spa1.4\spa3.5+4\spa1.3\spa4.5
               \right) \frac{L_0(-s_{12},-s_{35})}{s_{35}}
  -2\frac{\spa1.4^2\spa3.5^2\spa4.5\spb4.3\spb5.4}{\spa1.2\spa3.4^3} \frac{L_1(-s_{12},-s_{35})}{s_{35}^2} \nonumber\\
&&-\frac{\spa1.3\spa4.5\spb3.4}{\spa1.2\spa4.3^3} \left(2\spa1.3\spa4.5+4\spa1.4\spa3.5
               \right) \frac{L_0(-s_{12},-s_{45})}{s_{45}}
  -2\frac{\spa1.3^2\spa4.5^2\spa3.5\spb3.4\spb5.3}{\spa1.2\spa4.3^3} \frac{L_1(-s_{12},-s_{45})}{s_{45}^2} \nonumber\\
&&-\frac{2\spa3.5\spa4.5\spb2.5^2\spb3.4}{\spa3.4^3\spb1.2\spb3.5\spb4.5}
\end{eqnarray}
For the fermion loops the $g$-MHV and $\gamma$-MHV amplitudes are related in the same manner as the
leading order ones, 
\begin{equation}
\mathcal{A}^{N_f}(1_{\overline{q}}^-,2_q^+,3_{g}^-,4_{\gamma}^+,5_{\gamma}^+)
 = \mathcal{A}^{N_f}(1_{\overline{q}}^-,2_q^+,5_{g}^+,4_{\gamma}^+,3_{\gamma}^-) \;.
\end{equation}

\bibliography{Triphot}

\begin{thebibliography}{42}
\expandafter\ifx\csname natexlab\endcsname\relax\def\natexlab#1{#1}\fi
\expandafter\ifx\csname bibnamefont\endcsname\relax
  \def\bibnamefont#1{#1}\fi
\expandafter\ifx\csname bibfnamefont\endcsname\relax
  \def\bibfnamefont#1{#1}\fi
\expandafter\ifx\csname citenamefont\endcsname\relax
  \def\citenamefont#1{#1}\fi
\expandafter\ifx\csname url\endcsname\relax
  \def\url#1{\texttt{#1}}\fi
\expandafter\ifx\csname urlprefix\endcsname\relax\def\urlprefix{URL }\fi
\providecommand{\bibinfo}[2]{#2}
\providecommand{\eprint}[2][]{\url{#2}}

\bibitem[{\citenamefont{Bonvin et~al.}(1989)}]{Bonvin:1988yu}
\bibinfo{author}{\bibfnamefont{E.}~\bibnamefont{Bonvin}} \bibnamefont{et~al.}
  (\bibinfo{collaboration}{WA70 Collaboration}), \bibinfo{journal}{Z.Phys.}
  \textbf{\bibinfo{volume}{C41}}, \bibinfo{pages}{591} (\bibinfo{year}{1989}).

\bibitem[{\citenamefont{Albajar et~al.}(1988)}]{Albajar:1988im}
\bibinfo{author}{\bibfnamefont{C.}~\bibnamefont{Albajar}} \bibnamefont{et~al.}
  (\bibinfo{collaboration}{UA1 Collaboration}), \bibinfo{journal}{Phys.Lett.}
  \textbf{\bibinfo{volume}{B209}}, \bibinfo{pages}{385} (\bibinfo{year}{1988}).

\bibitem[{\citenamefont{Alitti et~al.}(1992)}]{Alitti:1992hn}
\bibinfo{author}{\bibfnamefont{J.}~\bibnamefont{Alitti}} \bibnamefont{et~al.}
  (\bibinfo{collaboration}{UA2 Collaboration}), \bibinfo{journal}{Phys.Lett.}
  \textbf{\bibinfo{volume}{B288}}, \bibinfo{pages}{386} (\bibinfo{year}{1992}).

\bibitem[{\citenamefont{Abe et~al.}(1993)}]{Abe:1992cy}
\bibinfo{author}{\bibfnamefont{F.}~\bibnamefont{Abe}} \bibnamefont{et~al.}
  (\bibinfo{collaboration}{CDF Collaboration}),
  \bibinfo{journal}{Phys.Rev.Lett.} \textbf{\bibinfo{volume}{70}},
  \bibinfo{pages}{2232} (\bibinfo{year}{1993}).

\bibitem[{\citenamefont{Abachi et~al.}(1996)}]{Abachi:1996qz}
\bibinfo{author}{\bibfnamefont{S.}~\bibnamefont{Abachi}} \bibnamefont{et~al.}
  (\bibinfo{collaboration}{D0 Collaboration}),
  \bibinfo{journal}{Phys.Rev.Lett.} \textbf{\bibinfo{volume}{77}},
  \bibinfo{pages}{5011} (\bibinfo{year}{1996}), \eprint{hep-ex/9603006}.

\bibitem[{\citenamefont{Aaltonen et~al.}(2011)}]{Aaltonen:2011vk}
\bibinfo{author}{\bibfnamefont{T.}~\bibnamefont{Aaltonen}} \bibnamefont{et~al.}
  (\bibinfo{collaboration}{CDF Collaboration}), \bibinfo{journal}{Phys.Rev.}
  \textbf{\bibinfo{volume}{D84}}, \bibinfo{pages}{052006}
  (\bibinfo{year}{2011}), \eprint{1106.5131}.

\bibitem[{\citenamefont{Abazov et~al.}(2010)}]{Abazov:2010ah}
\bibinfo{author}{\bibfnamefont{V.}~\bibnamefont{Abazov}} \bibnamefont{et~al.}
  (\bibinfo{collaboration}{D0 Collaboration}), \bibinfo{journal}{Phys.Lett.}
  \textbf{\bibinfo{volume}{B690}}, \bibinfo{pages}{108} (\bibinfo{year}{2010}),
  \eprint{1002.4917}.

\bibitem[{\citenamefont{Aaltonen et~al.}(2013)}]{Aaltonen:2012jd}
\bibinfo{author}{\bibfnamefont{T.}~\bibnamefont{Aaltonen}} \bibnamefont{et~al.}
  (\bibinfo{collaboration}{CDF Collaboration}),
  \bibinfo{journal}{Phys.Rev.Lett.} \textbf{\bibinfo{volume}{110}},
  \bibinfo{pages}{101801} (\bibinfo{year}{2013}), \eprint{1212.4204}.

\bibitem[{\citenamefont{Chatrchyan
  et~al.}(2012{\natexlab{a}})}]{Chatrchyan:2011qt}
\bibinfo{author}{\bibfnamefont{S.}~\bibnamefont{Chatrchyan}}
  \bibnamefont{et~al.} (\bibinfo{collaboration}{CMS Collaboration}),
  \bibinfo{journal}{JHEP} \textbf{\bibinfo{volume}{1201}}, \bibinfo{pages}{133}
  (\bibinfo{year}{2012}{\natexlab{a}}), \eprint{1110.6461}.

\bibitem[{\citenamefont{Aad et~al.}(2013{\natexlab{a}})}]{Aad:2012tba}
\bibinfo{author}{\bibfnamefont{G.}~\bibnamefont{Aad}} \bibnamefont{et~al.}
  (\bibinfo{collaboration}{ATLAS Collaboration}), \bibinfo{journal}{JHEP}
  \textbf{\bibinfo{volume}{1301}}, \bibinfo{pages}{086}
  (\bibinfo{year}{2013}{\natexlab{a}}), \eprint{1211.1913}.

\bibitem[{\citenamefont{Aad et~al.}(2013{\natexlab{b}})}]{Aad:2013zba}
\bibinfo{author}{\bibfnamefont{G.}~\bibnamefont{Aad}} \bibnamefont{et~al.}
  (\bibinfo{collaboration}{ATLAS Collaboration})
  (\bibinfo{year}{2013}{\natexlab{b}}), \eprint{1311.1440}.

\bibitem[{\citenamefont{Chatrchyan
  et~al.}(2013{\natexlab{a}})}]{Chatrchyan:2013mwa}
\bibinfo{author}{\bibfnamefont{S.}~\bibnamefont{Chatrchyan}}
  \bibnamefont{et~al.} (\bibinfo{collaboration}{CMS Collaboration})
  (\bibinfo{year}{2013}{\natexlab{a}}), \eprint{1311.6141}.

\bibitem[{\citenamefont{Aad et~al.}(2012)}]{Aad:2012tfa}
\bibinfo{author}{\bibfnamefont{G.}~\bibnamefont{Aad}} \bibnamefont{et~al.}
  (\bibinfo{collaboration}{ATLAS Collaboration}), \bibinfo{journal}{Phys.Lett.}
  \textbf{\bibinfo{volume}{B716}}, \bibinfo{pages}{1} (\bibinfo{year}{2012}),
  \eprint{1207.7214}.

\bibitem[{\citenamefont{Chatrchyan
  et~al.}(2012{\natexlab{b}})}]{Chatrchyan:2012ufa}
\bibinfo{author}{\bibfnamefont{S.}~\bibnamefont{Chatrchyan}}
  \bibnamefont{et~al.} (\bibinfo{collaboration}{CMS Collaboration}),
  \bibinfo{journal}{Phys.Lett.} \textbf{\bibinfo{volume}{B716}},
  \bibinfo{pages}{30} (\bibinfo{year}{2012}{\natexlab{b}}), \eprint{1207.7235}.

\bibitem[{\citenamefont{Catani et~al.}(2002)\citenamefont{Catani, Fontannaz,
  Guillet, and Pilon}}]{Catani:2002ny}
\bibinfo{author}{\bibfnamefont{S.}~\bibnamefont{Catani}},
  \bibinfo{author}{\bibfnamefont{M.}~\bibnamefont{Fontannaz}},
  \bibinfo{author}{\bibfnamefont{J.}~\bibnamefont{Guillet}}, \bibnamefont{and}
  \bibinfo{author}{\bibfnamefont{E.}~\bibnamefont{Pilon}},
  \bibinfo{journal}{JHEP} \textbf{\bibinfo{volume}{0205}}, \bibinfo{pages}{028}
  (\bibinfo{year}{2002}), \eprint{hep-ph/0204023}.

\bibitem[{\citenamefont{Bourhis et~al.}(1998)\citenamefont{Bourhis, Fontannaz,
  and Guillet}}]{Bourhis:1997yu}
\bibinfo{author}{\bibfnamefont{L.}~\bibnamefont{Bourhis}},
  \bibinfo{author}{\bibfnamefont{M.}~\bibnamefont{Fontannaz}},
  \bibnamefont{and} \bibinfo{author}{\bibfnamefont{J.}~\bibnamefont{Guillet}},
  \bibinfo{journal}{Eur.Phys.J.} \textbf{\bibinfo{volume}{C2}},
  \bibinfo{pages}{529} (\bibinfo{year}{1998}), \eprint{hep-ph/9704447}.

\bibitem[{\citenamefont{Gehrmann-De~Ridder and
  Glover}(1998)}]{GehrmannDeRidder:1997gf}
\bibinfo{author}{\bibfnamefont{A.}~\bibnamefont{Gehrmann-De~Ridder}}
  \bibnamefont{and} \bibinfo{author}{\bibfnamefont{E.~N.}
  \bibnamefont{Glover}}, \bibinfo{journal}{Nucl.Phys.}
  \textbf{\bibinfo{volume}{B517}}, \bibinfo{pages}{269} (\bibinfo{year}{1998}),
  \eprint{hep-ph/9707224}.

\bibitem[{\citenamefont{Frixione}(1998)}]{Frixione:1998jh}
\bibinfo{author}{\bibfnamefont{S.}~\bibnamefont{Frixione}},
  \bibinfo{journal}{Phys.Lett.} \textbf{\bibinfo{volume}{B429}},
  \bibinfo{pages}{369} (\bibinfo{year}{1998}), \eprint{hep-ph/9801442}.

\bibitem[{\citenamefont{Binoth et~al.}(2000)\citenamefont{Binoth, Guillet,
  Pilon, and Werlen}}]{Binoth:1999qq}
\bibinfo{author}{\bibfnamefont{T.}~\bibnamefont{Binoth}},
  \bibinfo{author}{\bibfnamefont{J.}~\bibnamefont{Guillet}},
  \bibinfo{author}{\bibfnamefont{E.}~\bibnamefont{Pilon}}, \bibnamefont{and}
  \bibinfo{author}{\bibfnamefont{M.}~\bibnamefont{Werlen}},
  \bibinfo{journal}{Eur.Phys.J.} \textbf{\bibinfo{volume}{C16}},
  \bibinfo{pages}{311} (\bibinfo{year}{2000}), \eprint{hep-ph/9911340}.

\bibitem[{\citenamefont{Catani et~al.}(2012)\citenamefont{Catani, Cieri,
  de~Florian, Ferrera, and Grazzini}}]{Catani:2011qz}
\bibinfo{author}{\bibfnamefont{S.}~\bibnamefont{Catani}},
  \bibinfo{author}{\bibfnamefont{L.}~\bibnamefont{Cieri}},
  \bibinfo{author}{\bibfnamefont{D.}~\bibnamefont{de~Florian}},
  \bibinfo{author}{\bibfnamefont{G.}~\bibnamefont{Ferrera}}, \bibnamefont{and}
  \bibinfo{author}{\bibfnamefont{M.}~\bibnamefont{Grazzini}},
  \bibinfo{journal}{Phys.Rev.Lett.} \textbf{\bibinfo{volume}{108}},
  \bibinfo{pages}{072001} (\bibinfo{year}{2012}), \eprint{1110.2375}.

\bibitem[{\citenamefont{Del~Duca et~al.}(2003)\citenamefont{Del~Duca, Maltoni,
  Nagy, and Trocsanyi}}]{DelDuca:2003uz}
\bibinfo{author}{\bibfnamefont{V.}~\bibnamefont{Del~Duca}},
  \bibinfo{author}{\bibfnamefont{F.}~\bibnamefont{Maltoni}},
  \bibinfo{author}{\bibfnamefont{Z.}~\bibnamefont{Nagy}}, \bibnamefont{and}
  \bibinfo{author}{\bibfnamefont{Z.}~\bibnamefont{Trocsanyi}},
  \bibinfo{journal}{JHEP} \textbf{\bibinfo{volume}{0304}}, \bibinfo{pages}{059}
  (\bibinfo{year}{2003}), \eprint{hep-ph/0303012}.

\bibitem[{\citenamefont{Gehrmann
  et~al.}(2013{\natexlab{a}})\citenamefont{Gehrmann, Greiner, and
  Heinrich}}]{Gehrmann:2013aga}
\bibinfo{author}{\bibfnamefont{T.}~\bibnamefont{Gehrmann}},
  \bibinfo{author}{\bibfnamefont{N.}~\bibnamefont{Greiner}}, \bibnamefont{and}
  \bibinfo{author}{\bibfnamefont{G.}~\bibnamefont{Heinrich}},
  \bibinfo{journal}{JHEP} \textbf{\bibinfo{volume}{1306}}, \bibinfo{pages}{058}
  (\bibinfo{year}{2013}{\natexlab{a}}), \eprint{1303.0824}.

\bibitem[{\citenamefont{Gehrmann
  et~al.}(2013{\natexlab{b}})\citenamefont{Gehrmann, Greiner, and
  Heinrich}}]{Gehrmann:2013bga}
\bibinfo{author}{\bibfnamefont{T.}~\bibnamefont{Gehrmann}},
  \bibinfo{author}{\bibfnamefont{N.}~\bibnamefont{Greiner}}, \bibnamefont{and}
  \bibinfo{author}{\bibfnamefont{G.}~\bibnamefont{Heinrich}}
  (\bibinfo{year}{2013}{\natexlab{b}}), \eprint{1308.3660}.

\bibitem[{\citenamefont{Bern et~al.}(2013)\citenamefont{Bern, Dixon, Cordero,
  Hoeche, Ita et~al.}}]{Bern:2013bha}
\bibinfo{author}{\bibfnamefont{Z.}~\bibnamefont{Bern}},
  \bibinfo{author}{\bibfnamefont{L.}~\bibnamefont{Dixon}},
  \bibinfo{author}{\bibfnamefont{F.~F.} \bibnamefont{Cordero}},
  \bibinfo{author}{\bibfnamefont{S.}~\bibnamefont{Hoeche}},
  \bibinfo{author}{\bibfnamefont{H.}~\bibnamefont{Ita}}, \bibnamefont{et~al.}
  (\bibinfo{year}{2013}), \eprint{1312.0592}.

\bibitem[{\citenamefont{Badger et~al.}(2013)\citenamefont{Badger, Guffanti, and
  Yundin}}]{Badger:2013ava}
\bibinfo{author}{\bibfnamefont{S.}~\bibnamefont{Badger}},
  \bibinfo{author}{\bibfnamefont{A.}~\bibnamefont{Guffanti}}, \bibnamefont{and}
  \bibinfo{author}{\bibfnamefont{V.}~\bibnamefont{Yundin}}
  (\bibinfo{year}{2013}), \eprint{1312.5927}.

\bibitem[{\citenamefont{Bern et~al.}(2014)\citenamefont{Bern, Dixon,
  Febres~Cordero, Hoeche, Ita et~al.}}]{Bern:2014vza}
\bibinfo{author}{\bibfnamefont{Z.}~\bibnamefont{Bern}},
  \bibinfo{author}{\bibfnamefont{L.}~\bibnamefont{Dixon}},
  \bibinfo{author}{\bibfnamefont{F.}~\bibnamefont{Febres~Cordero}},
  \bibinfo{author}{\bibfnamefont{S.}~\bibnamefont{Hoeche}},
  \bibinfo{author}{\bibfnamefont{H.}~\bibnamefont{Ita}}, \bibnamefont{et~al.}
  (\bibinfo{year}{2014}), \eprint{1402.4127}.

\bibitem[{\citenamefont{Aad
  et~al.}(2013{\natexlab{c}})}]{ATL-PHYS-PUB-2013-006}
\bibinfo{author}{\bibfnamefont{G.}~\bibnamefont{Aad}} \bibnamefont{et~al.}
  (\bibinfo{collaboration}{ATLAS Collaboration})
  (\bibinfo{year}{2013}{\natexlab{c}}), \eprint{ATL-PHYS-PUB-2013-006}.

\bibitem[{\citenamefont{Chatrchyan
  et~al.}(2013{\natexlab{b}})}]{CMS-PAS-SMP-13-009}
\bibinfo{author}{\bibfnamefont{S.}~\bibnamefont{Chatrchyan}}
  \bibnamefont{et~al.} (\bibinfo{collaboration}{CMS Collaboration})
  (\bibinfo{year}{2013}{\natexlab{b}}), \eprint{CMS-PAS-SMP-13-009}.

\bibitem[{\citenamefont{Bozzi et~al.}(2011)\citenamefont{Bozzi, Campanario,
  Rauch, and Zeppenfeld}}]{Bozzi:2011en}
\bibinfo{author}{\bibfnamefont{G.}~\bibnamefont{Bozzi}},
  \bibinfo{author}{\bibfnamefont{F.}~\bibnamefont{Campanario}},
  \bibinfo{author}{\bibfnamefont{M.}~\bibnamefont{Rauch}}, \bibnamefont{and}
  \bibinfo{author}{\bibfnamefont{D.}~\bibnamefont{Zeppenfeld}},
  \bibinfo{journal}{Phys.Rev.} \textbf{\bibinfo{volume}{D84}},
  \bibinfo{pages}{074028} (\bibinfo{year}{2011}), \eprint{1107.3149}.

\bibitem[{\citenamefont{Del~Duca et~al.}(2000)\citenamefont{Del~Duca, Kilgore,
  and Maltoni}}]{DelDuca:1999pa}
\bibinfo{author}{\bibfnamefont{V.}~\bibnamefont{Del~Duca}},
  \bibinfo{author}{\bibfnamefont{W.~B.} \bibnamefont{Kilgore}},
  \bibnamefont{and} \bibinfo{author}{\bibfnamefont{F.}~\bibnamefont{Maltoni}},
  \bibinfo{journal}{Nucl.Phys.} \textbf{\bibinfo{volume}{B566}},
  \bibinfo{pages}{252} (\bibinfo{year}{2000}), \eprint{hep-ph/9910253}.

\bibitem[{\citenamefont{Bern et~al.}(1995)\citenamefont{Bern, Dixon, and
  Kosower}}]{Bern:1994fz}
\bibinfo{author}{\bibfnamefont{Z.}~\bibnamefont{Bern}},
  \bibinfo{author}{\bibfnamefont{L.~J.} \bibnamefont{Dixon}}, \bibnamefont{and}
  \bibinfo{author}{\bibfnamefont{D.~A.} \bibnamefont{Kosower}},
  \bibinfo{journal}{Nucl.Phys.} \textbf{\bibinfo{volume}{B437}},
  \bibinfo{pages}{259} (\bibinfo{year}{1995}), \eprint{hep-ph/9409393}.

\bibitem[{\citenamefont{Britto et~al.}(2005)\citenamefont{Britto, Cachazo, and
  Feng}}]{Britto:2004nc}
\bibinfo{author}{\bibfnamefont{R.}~\bibnamefont{Britto}},
  \bibinfo{author}{\bibfnamefont{F.}~\bibnamefont{Cachazo}}, \bibnamefont{and}
  \bibinfo{author}{\bibfnamefont{B.}~\bibnamefont{Feng}},
  \bibinfo{journal}{Nucl.Phys.} \textbf{\bibinfo{volume}{B725}},
  \bibinfo{pages}{275} (\bibinfo{year}{2005}), \eprint{hep-th/0412103}.

\bibitem[{\citenamefont{Britto et~al.}(2006)\citenamefont{Britto, Feng, and
  Mastrolia}}]{Britto:2006sj}
\bibinfo{author}{\bibfnamefont{R.}~\bibnamefont{Britto}},
  \bibinfo{author}{\bibfnamefont{B.}~\bibnamefont{Feng}}, \bibnamefont{and}
  \bibinfo{author}{\bibfnamefont{P.}~\bibnamefont{Mastrolia}},
  \bibinfo{journal}{Phys.Rev.} \textbf{\bibinfo{volume}{D73}},
  \bibinfo{pages}{105004} (\bibinfo{year}{2006}), \eprint{hep-ph/0602178}.

\bibitem[{\citenamefont{Mastrolia}(2009)}]{Mastrolia:2009dr}
\bibinfo{author}{\bibfnamefont{P.}~\bibnamefont{Mastrolia}},
  \bibinfo{journal}{Phys.Lett.} \textbf{\bibinfo{volume}{B678}},
  \bibinfo{pages}{246} (\bibinfo{year}{2009}), \eprint{0905.2909}.

\bibitem[{\citenamefont{Badger}(2009)}]{Badger:2008cm}
\bibinfo{author}{\bibfnamefont{S.}~\bibnamefont{Badger}},
  \bibinfo{journal}{JHEP} \textbf{\bibinfo{volume}{0901}}, \bibinfo{pages}{049}
  (\bibinfo{year}{2009}), \eprint{0806.4600}.

\bibitem[{\citenamefont{Maitre and Mastrolia}(2008)}]{Maitre:2007jq}
\bibinfo{author}{\bibfnamefont{D.}~\bibnamefont{Maitre}} \bibnamefont{and}
  \bibinfo{author}{\bibfnamefont{P.}~\bibnamefont{Mastrolia}},
  \bibinfo{journal}{Comput.Phys.Commun.} \textbf{\bibinfo{volume}{179}},
  \bibinfo{pages}{501} (\bibinfo{year}{2008}), \eprint{0710.5559}.

\bibitem[{\citenamefont{Campbell and Ellis}(1999)}]{Campbell:1999ah}
\bibinfo{author}{\bibfnamefont{J.~M.} \bibnamefont{Campbell}} \bibnamefont{and}
  \bibinfo{author}{\bibfnamefont{R.~K.} \bibnamefont{Ellis}},
  \bibinfo{journal}{Phys.Rev.} \textbf{\bibinfo{volume}{D60}},
  \bibinfo{pages}{113006} (\bibinfo{year}{1999}), \eprint{hep-ph/9905386}.

\bibitem[{\citenamefont{Campbell et~al.}(2011)\citenamefont{Campbell, Ellis,
  and Williams}}]{Campbell:2011bn}
\bibinfo{author}{\bibfnamefont{J.~M.} \bibnamefont{Campbell}},
  \bibinfo{author}{\bibfnamefont{R.~K.} \bibnamefont{Ellis}}, \bibnamefont{and}
  \bibinfo{author}{\bibfnamefont{C.}~\bibnamefont{Williams}},
  \bibinfo{journal}{JHEP} \textbf{\bibinfo{volume}{1107}}, \bibinfo{pages}{018}
  (\bibinfo{year}{2011}), \eprint{1105.0020}.

\bibitem[{\citenamefont{Campbell et~al.}(2014)\citenamefont{Campbell, Ellis,
  and Williams}}]{MCFMweb}
\bibinfo{author}{\bibfnamefont{J.~M.} \bibnamefont{Campbell}},
  \bibinfo{author}{\bibfnamefont{R.~K.} \bibnamefont{Ellis}}, \bibnamefont{and}
  \bibinfo{author}{\bibfnamefont{C.}~\bibnamefont{Williams}}
  (\bibinfo{year}{2014}), \eprint{http://mcfm.fnal.gov}.

\bibitem[{\citenamefont{Catani and Seymour}(1997)}]{Catani:1996vz}
\bibinfo{author}{\bibfnamefont{S.}~\bibnamefont{Catani}} \bibnamefont{and}
  \bibinfo{author}{\bibfnamefont{M.}~\bibnamefont{Seymour}},
  \bibinfo{journal}{Nucl.Phys.} \textbf{\bibinfo{volume}{B485}},
  \bibinfo{pages}{291} (\bibinfo{year}{1997}), \eprint{hep-ph/9605323}.

\bibitem[{\citenamefont{Catani et~al.}(2013)\citenamefont{Catani, Fontannaz,
  Guillet, and Pilon}}]{Catani:2013oma}
\bibinfo{author}{\bibfnamefont{S.}~\bibnamefont{Catani}},
  \bibinfo{author}{\bibfnamefont{M.}~\bibnamefont{Fontannaz}},
  \bibinfo{author}{\bibfnamefont{J.~P.} \bibnamefont{Guillet}},
  \bibnamefont{and} \bibinfo{author}{\bibfnamefont{E.}~\bibnamefont{Pilon}},
  \bibinfo{journal}{JHEP} \textbf{\bibinfo{volume}{1309}}, \bibinfo{pages}{007}
  (\bibinfo{year}{2013}), \eprint{1306.6498}.

\bibitem[{\citenamefont{Lai et~al.}(2010)\citenamefont{Lai, Guzzi, Huston, Li,
  Nadolsky et~al.}}]{Lai:2010vv}
\bibinfo{author}{\bibfnamefont{H.-L.} \bibnamefont{Lai}},
  \bibinfo{author}{\bibfnamefont{M.}~\bibnamefont{Guzzi}},
  \bibinfo{author}{\bibfnamefont{J.}~\bibnamefont{Huston}},
  \bibinfo{author}{\bibfnamefont{Z.}~\bibnamefont{Li}},
  \bibinfo{author}{\bibfnamefont{P.~M.} \bibnamefont{Nadolsky}},
  \bibnamefont{et~al.}, \bibinfo{journal}{Phys.Rev.}
  \textbf{\bibinfo{volume}{D82}}, \bibinfo{pages}{074024}
  (\bibinfo{year}{2010}), \eprint{1007.2241}.

\end{thebibliography}
\end{document}